\documentclass{aa}
\pdfoutput=1          
\usepackage{graphicx}
\usepackage{amsmath,amssymb,amsfonts}
\usepackage{txfonts}
\usepackage{booktabs}
\usepackage{xcolor}
\usepackage{tikz}
\usetikzlibrary{arrows.meta, positioning, calc}
\usepackage[hidelinks]{hyperref}   
\usepackage{orcidlink}    

\AtBeginDocument{\nolinenumbers\let\linenumbers\relax}

\begin{document}

   \title{Graph reconstruction from random-walk co-visitation}
   \subtitle{Geometric, empirical, and controlled networks}

   \author{Marko Imbri\v{s}ak\thanks{\emph{marko.imbrisak@gmail.com}}\inst{1}\orcidlink{0000-0002-2773-8617}\hspace*{0.35em}
     \and Kre\v{s}imir Tisani\'{c}\inst{1}\orcidlink{0000-0001-6382-4937}}
   \authorrunning{M. Imbri\v{s}ak and K. Tisani\'c}

   \institute{
     Independent Researcher, Zagreb, Croatia
   }

   \date{}

  \abstract
   {
     Reconstructing an unknown graph from the trajectory of a random
     walk is a recurring problem, arising for spatial correlation
     networks in astrophysics and for connectivity inference in
     network science.
   }
   {
     We present a reconstruction pipeline whose observable is the
     random-walk co-visitation matrix, whose model is a pairwise
     edge-weight basis, and whose fitter is a frame-balanced
     Levenberg--Marquardt (fbLM) scheme with per-node group weights
     and a self-calibrated edge readout. We test it across networks of
     different structural character.
   }
   {
     We describe the procedure in full and apply it to an email
     communication subgraph, to Delaunay and Voronoi networks built
     from a COSMOS sky catalogue, and to two controlled $12$-vertex
     test graphs, one unicyclic and one a tree, under both
     analytic-noise and finite-walk regimes.
     Reconstructions are scored against the ground-truth adjacency by
     true/false positives and the Matthews correlation coefficient
     (MCC).
   }
   {
     The pipeline reconstructs all test-beds with high fidelity and
     scales to full graph size. On the finite-walk data it recovers the
     COSMOS Delaunay and Voronoi graphs at MCC $\gtrsim 0.98$ up to
     their full extent, $N=119$ and $N=223$, the whole graph rather
     than a cut-out of it, and
     the empirical \texttt{email-Eu-core} graph at $N=240$ ($417$ edges),
     where the residual misses are the edges the walk never traverses. On the full Delaunay graph the
     graphical-lasso reference returns MCC $0.540$ against $0.988$ for
     fbLM. Each reconstructed edge carries a Fisher-propagated
     uncertainty, and the residual misses are almost entirely confined to
     edges the walk never traverses.
   }
   {
     In the finite-walk regime the limiting factor is walk coverage
     rather than the fit: essentially every edge the walk visits is
     recovered, so reconstructibility is governed by the sampling of
     the graph rather than by the estimator.
   }


   \maketitle

   \section{Introduction}
   \label{sec:intro}

Recovering the structure of a network from an observed dynamical
process is a long-standing problem in network science. When the
process is a random walk the question takes a specific form: given
samples of a walker diffusing over an unknown graph, can the edge set
be recovered? It arises wherever the dynamics are accessible but the
wiring is not, as in the inference of interaction networks from time
series, of contact structures from spreading processes, or of spatial
adjacency from a diffusive tracer \citep{newman2018}. Several routes
are established. One treats the walk as a metric and inverts the
matrix of mean first-hitting times \citep{wittmann2009}; another
places a structural prior on the adjacency and infers it jointly with
the dynamics, so that the observations enter through a likelihood
rather than through a summary statistic \citep{peixoto2019}; a third
works from information theory, using the mutual information between a
random graph and a stochastic process running on it to quantify
reconstructability through uncertainty coefficients and to exhibit it as
dual to predictability \citep{murphy2024}. A random walk is one such
process, so that line of work asks what any estimator could in
principle extract from a trajectory, where we ask what one particular
estimator does extract. What
these approaches share is that the graph is abstract: its vertices
carry no coordinates, and a reconstruction can be scored only
combinatorially.

Two features of the problem deserve emphasis at the outset. The first
is \emph{reconstructibility}: a random walk can only carry information
about the edges it actually traverses, so any edge that the sampled
trajectory never crosses is unrecoverable in principle from that
sample, independently of the estimator. The practical quality of a
reconstruction is therefore bounded by how well the walk covers the
graph, a point we return to quantitatively in
Sect.~\ref{sec:results}. The second is \emph{identifiability}: the
walk observable must retain enough of the pairwise structure to
distinguish genuinely different graphs. By \emph{pairwise structure} we
mean information carried by a pair of vertices and not reconstructible
from per-vertex quantities alone: which pairs are joined, as against how
many neighbours each vertex has. Two graphs with the same degree
sequence agree in every node-level summary and can still disagree on
every edge, so an observable that collapses to such a summary, the
stationary degree distribution say, cannot separate them. The pipeline used here is
built specifically to avoid that collapse, using a pairwise
observable and a pairwise model, as summarised in
Sect.~\ref{sec:method}.

Geometric graphs built from point sets are a demanding test-bed for
such methods, and they are of independent interest in astrophysics,
where the large-scale galaxy distribution is itself routinely treated
as a network. \citet{hong2015} measured degree, closeness and
betweenness centrality on a graph built from the COSMOS catalogue and
used them to separate void, wall and cluster populations;
\citet{deregt2018} carried the same analysis across redshift slices
of that field. \citet{coutinho2016} compared seven rules for
assigning a network to a galaxy distribution and found that spatial
proximity alone gives the closest correspondence with the physical
properties of the connected galaxies. At the level of the filamentary
skeleton, the connectivity of a node, the number of filaments meeting
at it, depends on halo mass and on cosmology \citep{codis2018}.

In all of this work the graph is \emph{constructed} rather than
inferred. A linking length, a nearest-neighbour rule or a tessellation
turns positions into edges, and the edge set is a modelling choice
whose consequences are then explored; that seven such rules had to be
compared \citep{coutinho2016} is a symptom of the same thing. We are
not aware of an astrophysical analysis in which the graph is instead
the object of inference, recovered from an observed dynamical process
running on it. The geometric ingredients are nevertheless familiar
ones: the Delaunay triangulation and its dual Voronoi tessellation
encode the local adjacency and the local density field of a point
distribution, the Delaunay Tessellation Field Estimator uses that
structure to reconstruct continuous fields from discrete galaxy
positions \citep{vandeweygaert1994}, and the Voronoi tessellation
provides a stochastic-geometry model of the large-scale distribution
\citep{icke1987}. This paper does not close the gap on real
observations, since the walks used here are simulated on a known
graph. What it does is establish the estimator, its coverage limit and
its uncertainty structure on geometric graphs of exactly the kind that
literature builds, drawn from a COSMOS field. Because the vertices
carry true sky coordinates, a reconstruction can be inspected directly
against the ground truth on the plane of the sky, which an abstract
benchmark graph does not allow.

This paper is organised as follows.
Section~\ref{sec:method} describes the reconstruction procedure in
full: the co-visitation observable, the pairwise model, the fbLM
fitter, the gauge fixing, and the per-node readout, summarised in the
schematic of Fig.~\ref{fig:schematic}.
Section~\ref{sec:datasets} defines the datasets and controlled test
graphs used, together with the numerical setup for the noise and
walk-sampling regimes.
Section~\ref{sec:results} reports the reconstructions and the
coverage-limited cost curve.
Section~\ref{sec:discussion} discusses the results, the role of
reconstructibility, and directions toward larger and adversarial
settings.

   \section{Method}
   \label{sec:method}

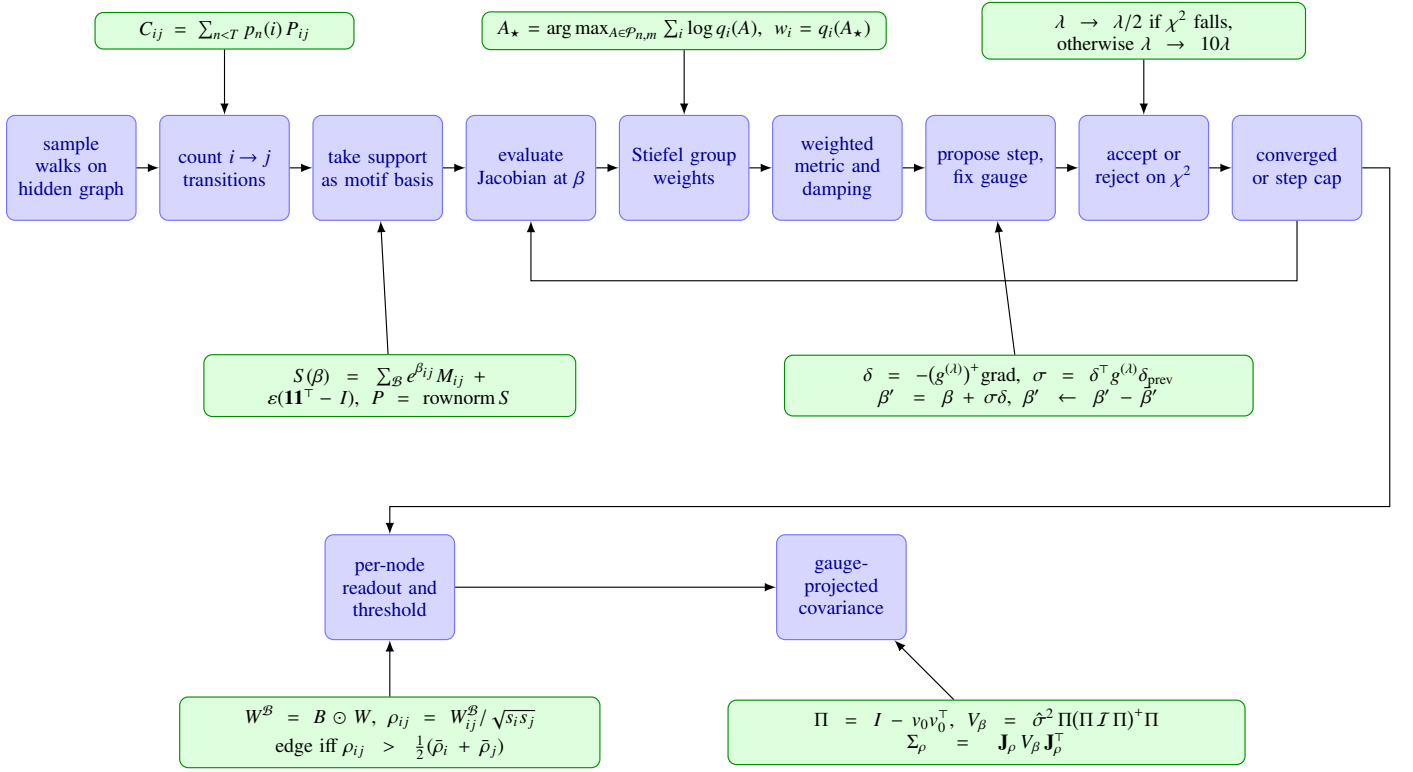
\begin{figure*}[tp]
\centering
\resizebox{\textwidth}{!}{%
\begin{tikzpicture}[
  x=1mm, y=1mm, font=\scriptsize,
  proc/.style={rectangle, rounded corners=3pt, draw=blue!55, fill=blue!16,
               text=blue!58!black, align=center, text width=15mm,
               minimum height=13mm, inner sep=1.5pt, line width=0.4pt},
  eqbox/.style={rectangle, rounded corners=3pt, draw=green!55!black,
                fill=green!13, align=center, inner sep=3pt, line width=0.4pt},
  fl/.style={-latex, line width=0.4pt}
]

\node[proc] (walk) at ( 8.5,0) {sample walks on hidden graph};
\node[proc] (cnt)  at (27.5,0) {count $i\!\to\!j$ transitions};
\node[proc] (supp) at (46.5,0) {take support as motif basis};
\node[proc] (jac)  at (65.5,0) {evaluate Jacobian at $\beta$};
\node[proc] (wts)  at (84.5,0) {Stiefel group weights};
\node[proc] (metr) at (103.5,0) {weighted metric and damping};
\node[proc] (prop) at (122.5,0) {propose step, fix gauge};
\node[proc] (acc)  at (141.5,0) {accept or reject on $\chi^{2}$};
\node[proc] (conv) at (160.5,0) {converged or step cap};

\foreach \a/\b in {walk/cnt, cnt/supp, supp/jac, jac/wts, wts/metr,
                   metr/prop, prop/acc, acc/conv}
  \draw[fl] (\a) -- (\b);

\node[eqbox, text width=30mm] (eqC) at ( 27.5,17)
  {$C_{ij}=\sum_{n<T} p_n(i)\,P_{ij}$};
\node[eqbox, text width=48mm] (eqW) at ( 84.5,17)
  {$A_\star=\arg\max_{A\in\mathcal P_{n,m}}\sum_i\log q_i(A)$,\;
   $w_i=q_i(A_\star)$};
\node[eqbox, text width=38mm] (eqA) at (141.5,17)
  {$\lambda\to\lambda/2$ if $\chi^{2}$ falls,\\ otherwise $\lambda\to10\lambda$};

\draw[fl] (eqC) -- (cnt);
\draw[fl] (eqW) -- (wts);
\draw[fl] (eqA) -- (acc);

\node[eqbox, text width=44mm] (eqD) at ( 48,-27)
  {$S(\beta)=\sum_{\mathcal B} e^{\beta_{ij}}M_{ij}+\varepsilon(\mathbf 1\mathbf 1^{\top}\!-I)$,\;
   $P=\mathrm{rownorm}\,S$};
\node[eqbox, text width=56mm] (eqP) at (126,-27)
  {$\delta=-\big(g^{(\lambda)}\big)^{+}\mathrm{grad}$,\;
   $\sigma=\delta^{\top}g^{(\lambda)}\delta_{\mathrm{prev}}$\\
   $\beta'=\beta+\sigma\delta$,\;
   $\beta'\leftarrow\beta'-\bar\beta'$};

\draw[fl] (eqD) -- (supp);
\draw[fl] (eqP) -- (prop);

\draw[fl] (conv.south) -- (160.5,-14) -- (65.5,-14) -- (jac.south);

\node[proc] (read) at ( 48,-52) {per-node readout and threshold};
\node[proc] (unc)  at (104,-52) {gauge-projected covariance};

\draw[fl] (conv.east) -- (172,0) -- (172,-42) -- (48,-42) -- (read.north);
\draw[fl] (read) -- (unc);

\node[eqbox, text width=50mm] (eqR) at ( 48,-70)
  {$W^{\mathcal B}=B\odot W$,\;
   $\rho_{ij}=W^{\mathcal B}_{ij}/\sqrt{s_i s_j}$\\
   edge iff $\rho_{ij}>\tfrac12(\bar\rho_i+\bar\rho_j)$};
\node[eqbox, text width=62mm] (eqU) at (122,-70)
  {$\Pi=I-v_0v_0^{\top}$,\;
   $V_\beta=\hat\sigma^{2}\,\Pi\big(\Pi\,\mathcal I\,\Pi\big)^{+}\Pi$\\
   $\Sigma_\rho=\mathbf J_\rho\,V_\beta\,\mathbf J_\rho^{\top}$};

\draw[fl] (eqR) -- (read);
\draw[fl] (eqU) -- (unc);

\end{tikzpicture}%
}
\caption{
   Flow of the reconstruction procedure. Blue boxes are the steps in
   order; green boxes carry the equations each step applies. The upper
   chain runs from the sampled walks to a converged fit: transitions are
   counted into the co-visitation $C$, its support fixes the pairwise
   motif basis, and the fbLM iteration (Sect.~\ref{sec:dynwlm}) then
   cycles over the Jacobian, the Stiefel group weights, the damped
   weighted metric, the proposal, and the acceptance ladder, returning to
   a fresh Jacobian until convergence. Two features distinguish the
   proposal step: the unnormalised geometric scale $\sigma$, which
   self-brakes on overshoot, and the gauge fix
   $\beta'\leftarrow\beta'-\bar\beta'$, which removes the flat direction
   left by the invariance of $C$ under $\beta\to\beta+c\mathbf 1$. The
   converged weights feed two independent outputs: the per-vertex readout
   (Sect.~\ref{sec:readout}), which produces the edge set with no absolute
   threshold, and the gauge-projected covariance
   (Sect.~\ref{sec:uq}), which attaches an uncertainty to each edge. The
   same projector $\Pi$ that fixes the gauge in the fit reappears there as
   the restriction of the Fisher metric to the quotient. The ground-truth
   adjacency $A$ enters only in the final diagnostic scoring, never in the
   fit.
}
\label{fig:schematic}
\end{figure*}

The full pipeline is summarised in Fig.~\ref{fig:schematic}. It maps
a vector of pairwise log-weights $\beta$ to a reconstructed edge set
$\hat E$ through a forward model, an iterative fit, and a readout.
Both the observable and the model are chosen against a specific
failure, and since avoiding those two failures is what the pipeline is
for, we state them before the procedure.

The first belongs to the observable. If the walk is summarised by its
marginal occupation, then in $p_{n'}(j)=\sum_i p_{n'-1}(i)P_{ij}$ the
sum over the previous vertex has already been taken and the pair
$(i,j)$ is no longer present in the record. On a connected,
non-bipartite graph the marginal also approaches its stationary value
$\pi_i\propto k_i$, with $k_i$ the degree of vertex $i$, within a few
steps, so what such a record carries is
essentially the degree sequence: graphs sharing a degree sequence are
indistinguishable in it, and the most a fit can recover from it is a
degree-driven null model. The co-visitation of Eq.~\eqref{eq:covis} is
chosen because it records the pair before the row sum is taken.

The second belongs to the model. A node-potential model gives each
vertex a single parameter $u_i$ and builds the weight of a pair
additively from its two endpoints, which after normalisation takes the
form $S_{ij}=Z-u_i-u_j$ with $Z$ a constant. It has $n$ parameters
rather than
$\binom{n}{2}$, and the weight of a pair is fixed once the potentials
of its endpoints are, so two pairs whose endpoints carry equal
potentials cannot be told apart whatever the data say. Clustered
structure, where the presence of an edge depends on the pair and not on
its endpoints separately, lies outside the reach of such a model.
Equation~\eqref{eq:S} therefore carries one parameter per candidate
pair.

Neither change is any use alone: a pairwise model fitted to a marginal
observable has no pairwise information to fit, and a pairwise
observable fitted with a node-potential model has nowhere to put what
it carries. What follows is the operational procedure, with the weight
solver given in Appendix~\ref{app:stiefel}.

\subsection{Forward model}
\label{sec:forward}

Let the graph have $n$ vertices, each candidate pair carrying a
log-weight $\beta_{ij}$.
Let $\mathcal B\subseteq\{(i,j):i<j\}$ denote the candidate basis and
$m=|\mathcal B|$. In the finite-walk regime $\mathcal B$ is the observed
support, $\mathcal B=\{(i,j):\hat C_{ij}+\hat C_{ji}>0\}$, the directed
counts being symmetrised before the test; in the multiplicative-noise
regime of Sect.~\ref{sec:setup} every pair carries a parameter and
$\mathcal B=\{(i,j):i<j\}$, so that $m=\binom{n}{2}$. The coverage
experiment of Sect.~\ref{sec:res-cost} is the one deliberate exception:
its observations are sampled walks, but the full pairwise basis is
retained so that the fitted scores can be ranked over edges and
non-edges alike. The model is
\begin{equation}
   S(\beta)=\sum_{(i,j)\in\mathcal B} e^{\beta_{ij}}\,M_{ij}
            +\varepsilon\,(\mathbf 1\mathbf 1^{\top}-I),
   \label{eq:S}
\end{equation}
where $M_{ij}$ is the indicator matrix of pair $(i,j)$ and
$\varepsilon=10^{-12}$ is a floor that keeps the row normalisation well
defined for a vertex carrying no basis edge. Row normalisation,
$\mathrm{rownorm}(S)_{ij}=S_{ij}/\sum_k S_{ik}$, divides every row by
its own sum and so turns $S$ into the walk transition matrix
$P=\mathrm{rownorm}(S)$, whose rows sum to one and whose entry $P_{ij}$
is the probability that a walker standing at $i$ steps to $j$. The
observable is the \emph{co-visitation} matrix,
\begin{equation}
   C_{ij}=\sum_{n'=0}^{T-1} p_{n'}(i)\,P_{ij},
   \qquad p_{n'+1}=P^{\top}p_{n'},
   \label{eq:covis}
\end{equation}
with a prior $p_0$ over vertices and a walk length $T$. Unlike the
marginal occupation $p_{n'}$, the co-visitation retains the ordered
pair $(i,j)$ before the row-sum, and it is this retention that keeps
the observable sensitive to pairwise structure. From finite
trajectories $C$ is estimated empirically by counting observed
transitions $i\to j$.

\subsection{The frame-balanced fit}
\label{sec:dynwlm}

The parameters $\beta$ are fitted to the measured co-visitation matrix
$C^{\mathrm{obs}}$, the superscript separating the data from the model
prediction $C(\beta)$ of Eq.~\eqref{eq:covis}. It is the $\hat C$ of
Sect.~\ref{sec:setup}, obtained either by counting transitions in
sampled walks or by perturbing the analytic $C$, according to the
regime. The fit minimises $\chi^2=\lVert r\rVert^2$ with residual
$r=C(\beta)-C^{\mathrm{obs}}$, using a frame-balanced
Levenberg--Marquardt iteration. The distinguishing feature of the
scheme is that the residual is grouped \emph{by vertex}: we write
$r^{(i)}=C(\beta)_{i\cdot}-C^{\mathrm{obs}}_{i\cdot}\in\mathbb{R}^{n}$
for the residual on row $i$ of the co-visitation (the co-visitation
``seen from vertex $i$''), and
\begin{equation}
   J^{(i)}=\frac{\partial C_{i\cdot}}{\partial\beta}\in\mathbb{R}^{m\times n}
   \label{eq:Jblock}
\end{equation}
for the corresponding block of the Jacobian, i.e.\ the derivative of
vertex $i$'s co-visitation row with respect to all $m$ parameters,
evaluated by finite differences. The per-vertex Gauss--Newton block
is $g^{(0)}_i=J^{(i)}(J^{(i)})^{\top}$, an $m\times m$ matrix.

Each iteration proceeds as follows. Positive group weights $w_i$ are
obtained by optimising over an orthonormal frame. The Stiefel manifold
is the set of $n\times m$ real matrices whose rows are orthonormal,
equivalently the set of ordered $n$-tuples of mutually orthogonal unit
vectors in $\mathbb R^m$; a point of it is such a frame, and the
manifold is what is left of $\mathbb R^{n\times m}$ once orthonormality
is imposed. Assigning the $i$th row to vertex $i$, we take
\begin{equation}
   A_\star=\arg\max_{A\in\mathcal P_{n,m}}\;\sum_i \log q_i(A),
   \qquad
   w_i=q_i(A_\star),
   \label{eq:stiefel}
\end{equation}
where $\tilde g^{(0)}_i$ is the whitened per-vertex block, $a_i^{\top}$
are the rows of $A\in\mathbb R^{n\times m}$ and
$q_i(A)=a_i^{\top}\tilde g^{(0)}_i a_i+\epsilon$, the ridge sitting inside
the logarithm. The feasible set is the
rectangular polar factor,
\begin{equation}
   \mathcal P_{n,m}=
   \begin{cases}
      \{A: AA^{\top}=I_n\}, & n\le m,\\[2pt]
      \{A: A^{\top}A=I_m\}, & n\ge m,
   \end{cases}
   \label{eq:polarset}
\end{equation}
the first case covering every test-bed except the radialness control,
where $n=12$ exceeds $m=11$ and twelve orthonormal rows cannot exist.
The weights are thus the value of the objective's per-vertex term at
the optimum rather than the optimiser itself, and the construction is
given in full in Appendix~\ref{app:stiefel}.

Two properties of Eq.~\eqref{eq:stiefel} fix what those weights mean,
and they are what the name of the scheme records. It is
\emph{balanced} because the objective is a sum of logarithms, that is
the logarithm of $\prod_i q_i$: maximising it maximises the geometric
mean of the per-vertex terms, and a term approaching zero costs an
unbounded amount, so no vertex can be left without weight in order to
serve the others. The whitening of Appendix~\ref{app:stiefel} is what
makes that possible, since without it the largest block dominates and
the weights collapse onto the highest-degree vertex. It is a
\emph{frame} because orthonormality couples the vertices to one
another: the rows compete for mutually orthogonal directions, so $q_i$
measures not how much information vertex $i$ carries outright but how
much of it lies along the direction the frame has left available to it.
A vertex whose residual block is aligned with its own direction
receives a large weight, one whose information is already accounted for
along its neighbours' directions a small one. We accordingly call the
scheme \emph{frame-balanced} Levenberg--Marquardt (fbLM). The weighted metric
and gradient are
\begin{equation}
   g=\sum_i w_i\,g^{(0)}_i,
   \qquad
   \mathrm{grad}=\sum_i w_i\,J^{(i)}\,r^{(i)}.
   \label{eq:metric}
\end{equation}
Damping is applied through the mean of the metric diagonal,
\begin{equation}
   g^{(\lambda)}=g+\lambda\,\bar d(g)\,I,
   \qquad
   \bar d(g)=\frac{1}{m}\operatorname{tr}(g)=\frac{1}{m}\sum_{k} g_{kk},
   \label{eq:damp}
\end{equation}
so that the damping term scales with the typical curvature rather
than with an absolute constant; $\bar d(g)$ is simply the mean of the
diagonal entries of $g$. The step is
\begin{equation}
   \delta=-\,(g^{(\lambda)})^{+}\,\mathrm{grad},
   \label{eq:step}
\end{equation}
where $(\cdot)^{+}$ denotes the Moore--Penrose pseudoinverse.

Two features set fbLM apart from a standard Levenberg--Marquardt
step. First, successive steps are combined through an
\emph{unnormalised geometric scale}
\begin{equation}
   \sigma=\delta^{\top}\,g^{(\lambda)}\,\delta_{\mathrm{prev}},
   \qquad
   \beta'=\beta+\sigma\,\delta,
   \label{eq:scale}
\end{equation}
where $\delta_{\mathrm{prev}}$ is the previous accepted step. Because
$\sigma$ is not normalised to a cosine, it carries the curvature
scale of the problem: it shortens the update when consecutive
proposals disagree and reverses sign on overshoot, providing
self-braking without an explicit trust region. Second, the
co-visitation of the idealised model depends only on the row-normalised
$P$ and is therefore invariant under a common shift
$\beta\to\beta+c\,\mathbf 1$; to remove
this gauge freedom, which otherwise lets the pseudoinverse in
Eq.~\eqref{eq:step} drift along the null direction and overflow
$e^{\beta}$, we fix the gauge after each proposal,
\begin{equation}
   \beta'\leftarrow\beta'-\bar\beta',
   \qquad
   \bar\beta'=\frac1m\sum_k\beta'_k .
   \label{eq:gauge}
\end{equation}
The invariance is exact for the idealised model without the floor, and
it fixes the null direction uniquely. The floor of Eq.~\eqref{eq:S} does not scale with
$e^{c}$, so the identity $S(\beta+c\mathbf1)=e^{c}S(\beta)$ holds for the
implemented matrix only while $e^{\beta}\gg\varepsilon$; at the fitted
parameters that margin is some twelve decades and the derivative
statement below holds to machine precision, but the invariance is not a
global identity of the regularised model. Both $P$ and the readout $\rho$ of
Sect.~\ref{sec:readout} are ratios of terms of equal order in $S$, hence
homogeneous of degree zero in $S$; Euler's theorem for such a function
$F$ gives $\sum_{ij}S_{ij}\,\partial F/\partial S_{ij}=0$, and since
$\partial/\partial\beta_{ij}=S_{ij}\,\partial/\partial S_{ij}$ in the
logarithmic coordinates this reads
\begin{equation}
   \mathbf J^{\top}\mathbf 1=0,
   \qquad\text{hence}\qquad
   \mathbf J\mathbf J^{\top}\,\mathbf 1=0 .
   \label{eq:null}
\end{equation}
For the idealised model the all-ones vector is therefore an exact null
vector of the Fisher matrix. The floor breaks the identity globally,
and it also breaks the chain rule used above, since
$\partial S_{ij}/\partial\beta_{ij}=e^{\beta_{ij}}$ rather than
$S_{ij}$; at the fitted parameters, however, the parametrised weights
exceed $\varepsilon$ by some twelve orders of magnitude, so the same
direction remains numerically near-null and is cleanly separated from
the physical spectrum (Fig.~\ref{fig:gauge-spectrum}). We therefore
remove that direction explicitly rather than let its finite-difference
shadow be inverted.

The geometry behind this is simple. Adding the same constant $c$ to
every $\beta_{ij}$ multiplies $S$ by $e^{c}$ and leaves $P$, and with it
$C$, unchanged, so every point of the line
$\{\beta+c\mathbf 1: c\in\mathbb R\}$ yields exactly the same
prediction. That line is the \emph{orbit} of $\beta$ under the shift:
the set of parameter vectors the shift can carry it to, and along which
every observable is constant. What the data determine is therefore not
$\beta$ but the orbit it lies on, and the parameter space is the set of
orbits, written $\mathbb{R}^m/\mathbb{R}\mathbf 1$.
Equation~\eqref{eq:gauge} picks one point out of each orbit, the one
whose entries sum to zero; the choice makes the fit well posed and
changes nothing it predicts.

There is one such direction here, and only one, because the pairs
carrying parameters form a single connected component in every run
reported; a basis falling into $c$ components would allow an
independent constant per component and $c$ directions. Vertices the
walk never visits, of which the $N=240$ email graph has ten, carry no
parameter and so contribute none. The same degeneracy returns when the
metric has to be inverted rather than merely fixed, which is
Sect.~\ref{sec:uq}.
Acceptance uses the usual ladder: the proposal is accepted and
$\lambda\to\lambda/2$ if $\chi^2$ decreases, otherwise it is rejected
and $\lambda\to10\lambda$. Empirically the edge ranking stabilises
well before $\chi^2$ has fully converged, which makes the readout
below robust under noise.

\subsection{Readout}
\label{sec:readout}

Given the fitted weights $W=S(\beta_{\mathrm{fit}})$, let $B$ be the
indicator of the basis, $B_{ij}=1$ for $(i,j)\in\mathcal B$ and zero
otherwise, and restrict the weights to it, $W^{\mathcal B}=B\odot W$. We
then form vertex strengths $s_i=\sum_k W^{\mathcal B}_{ik}$ and the
self-calibrated coupling
\begin{equation}
   \rho_{ij}=
   \begin{cases}
      \dfrac{W^{\mathcal B}_{ij}}{\sqrt{s_i\,s_j}},
        & (i,j)\in\mathcal B \text{ and } s_is_j>0,\\[8pt]
      0, & \text{otherwise,}
   \end{cases}
   \label{eq:rho}
\end{equation}
Writing $\bar\rho_i=(n-1)^{-1}\sum_{j\neq i}\rho_{ij}$ for the mean
coupling at vertex $i$, an edge is declared when the coupling exceeds
the mean of its two endpoints,
\begin{equation}
   (i,j)\in\hat E
   \;\Longleftrightarrow\;
   (i,j)\in\mathcal B
   \;\text{and}\;
   \rho_{ij}>\tfrac12\big(\bar\rho_i+\bar\rho_j\big).
   \label{eq:readout}
\end{equation}
The second branch covers vertices the walk never visited, which after
masking have $s_i=0$; it keeps $\bar\rho_i$ well defined for them and is
implemented by evaluating the ratio only on basis pairs.
Masking before the strengths rather than after the comparison matters
for the same reason the floor exists: without it $s_i$ and $\bar\rho_i$
inherit contributions from pairs the walk never traversed.
The comparison is made over the motif basis, so the candidate set is
exactly the set of pairs the walk was observed to traverse. The
restriction is not cosmetic. Equation~\eqref{eq:S} carries a floor of
$10^{-12}$ on every off-diagonal entry, without which the row
normalisation of $S$ would be undefined for a vertex that has no basis
edge at all. For a vertex the walk never visited, that floor is the
whole row, so its $s_i$ is set by the floor itself; two such vertices
then see each other at $\rho\sim4\times10^{-3}$ against a threshold of
$\sim\!2\times10^{-4}$, and a set of $k$ never-visited vertices condenses
into a clique of $k(k-1)/2$ edges built out of nothing. Evaluating the
readout on the basis keeps a numerical guard belonging to the fit out
of the decision stage.

Two properties of Eq.~\eqref{eq:readout} are worth making explicit.

There is no absolute threshold, in particular no $1/N$ cut. A fixed
numerical cut would have to be calibrated against the size and the
density of the graph and against the overall normalisation of $W$, none
of which is known in advance. Here the comparison is against a quantity
computed from the same fitted weights at the two endpoints of the pair,
so it rescales with the graph of its own accord. Since $\rho$ is a
ratio of terms of equal order in $S$ it is moreover invariant under the
shift of Eq.~\eqref{eq:gauge}, and the decision inherits that
invariance, so the edge set cannot depend on a choice the data do not
fix.

The ground-truth adjacency $A$ enters neither the fit nor the readout.
It is available at all only because the test-beds are constructed, and
were any part of the pipeline to use it, by tuning the threshold to
maximise a score, by restricting the basis to true edges, or by
stopping the iteration where the score peaks, the numbers of
Sect.~\ref{sec:results} would be circular and the procedure could not
be run on a graph whose edges are unknown. Keeping $A$ out means that
nothing has to change when the answer is unavailable; it appears only
in the scoring.

\subsection{Uncertainty propagation}
\label{sec:uq}

Both uncertainty products reported below, the band on the fitted
observable and the per-edge uncertainty on the readout, are pushforwards
of one parameter covariance $V_\beta$, so it is worth being explicit about
what that covariance is. For the idealised model the Fisher matrix
$\mathcal I=\mathbf J\mathbf J^{\top}$ annihilates $\mathbf 1$ exactly
(Eq.~\ref{eq:null}), and the metric that carries information lives on the
quotient $\mathbb{R}^m/\mathbb{R}\mathbf 1$, whose inverse is what
$V_\beta$ must express in the ambient coordinates. In the regularised
implementation the same direction survives as a numerically near-null
shadow: the matrix is then formally invertible, but inverting it means
inverting that shadow, which is precisely what we avoid by projecting it
out.
Writing $v_0=\mathbf 1/\sqrt m$ for the unit null direction and
\begin{equation}
   \Pi = I - v_0 v_0^{\top} = I - \tfrac1m\mathbf 1\mathbf 1^{\top}
   \label{eq:proj}
\end{equation}
for the orthogonal projector onto $\{v:\sum_k v_k=0\}$, we take
\begin{equation}
   V_\beta=\hat\sigma^{2}\,\Pi\,\big(\Pi\,\mathcal I\,\Pi\big)^{+}\,\Pi,
   \qquad
   \hat\sigma^{2}=\frac{\chi^{2}}{N_{\mathrm{d}}-m+1},
   \label{eq:vbeta}
\end{equation}
with $N_{\mathrm{d}}$ the number of informative entries of the residual
matrix; the $+1$ restores the direction removed by the gauge, which is
not an identifiable parameter. Note that $\Pi v=v-\bar v\,\mathbf 1$ is the same mean-centring
operation as the gauge fix of Eq.~\eqref{eq:gauge}: the fit centres the
parameter, Eq.~\eqref{eq:vbeta} centres the directions, so the two
gauge choices agree. The complement is taken to be Euclidean because
$v_0$ lies in the kernel of $\mathcal I$, so Fisher-orthogonality does not
single out a complement at all.

\begin{figure*}
\centering
\includegraphics[width=0.92\textwidth]{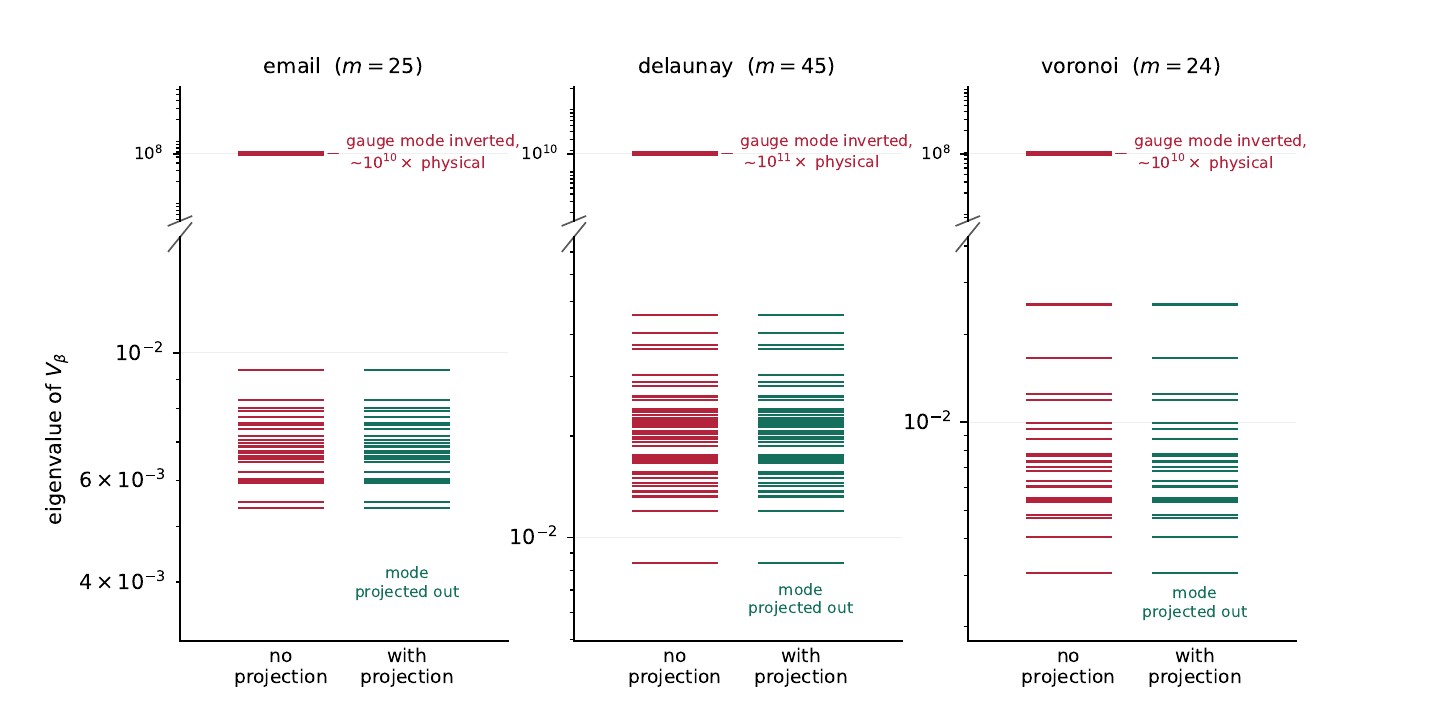}
\caption{
   Eigenvalue spectrum of the parameter covariance $V_\beta$ at $N=20$,
   without the gauge projection (left column of each panel) and with it
   (right column), drawn as a level diagram on a broken logarithmic axis.
   The physical levels are identical in the two columns: projecting does
   not move them. The only difference is the single mode in the upper
   sub-panel, produced when a plain pseudoinverse inverts the gauge
   direction instead of discarding it; it exceeds the largest physical
   eigenvalue by factors of $6\times10^{10}$
   (\texttt{email-Eu-core}), $3.9\times10^{11}$ (Delaunay) and
   $3.4\times10^{10}$ (Voronoi). Because the readout $\rho$ is a ratio and
   therefore only weakly dependent on $\beta$, this one mode is enough to
   dominate $\sigma(\rho)$ if it is left in.
}
\label{fig:gauge-spectrum}
\end{figure*}

The projector is not cosmetic. Analytically the null eigenvalue is zero,
but $\mathbf J$ is evaluated by finite differences with step $\Delta$, and
the leakage $\lVert\mathbf J^{\top}v_0\rVert/\lVert\mathbf J\rVert$ is
$\mathcal O(\Delta)$, placing the smallest eigenvalue at
$\mathcal O(\Delta^{2})$ rather than at zero. For our step
$\Delta=10^{-3}$ this puts it near $10^{-12}$, which is still above the
default relative cutoff of a Moore--Penrose pseudoinverse
($\sim\!10^{-15}$): a plain $\mathcal I^{+}$ therefore treats the gauge
direction as a genuine soft mode and inverts it, amplifying it by some
twelve orders of magnitude. We verified the scaling directly by varying
$\Delta$ over four decades: the leakage falls linearly and the smallest
eigenvalue quadratically, with the next eigenvalue unmoved, confirming
that the near-zero mode is a numerical shadow of a structural zero rather
than a weakly-determined physical direction. The spectrum is
correspondingly bimodal: at $N=20$ the null eigenvalue sits ten orders
of magnitude below the smallest physical one, so the cutoff has a wide
margin on both sides. Figure~\ref{fig:gauge-spectrum} shows the effect on
the covariance itself: without $\Pi$ the inverted gauge direction appears
as a single mode ten to eleven orders of magnitude above the entire
physical spectrum, while the physical levels themselves are untouched by
the projection.

The consequence differs between the two pushforwards, which is why the
distinction surfaced only in the per-edge product. The observable $C$ is
gauge-invariant and strongly $\beta$-dependent, so its Jacobian is
insensitive to the amplified direction and the band of
Sect.~\ref{sec:res-fit} is unchanged to within $0.05\%$ whether or not
$\Pi$ is applied. The readout $\rho$ is likewise gauge-invariant but, being
a ratio, is far less sensitive to $\beta$; the same absolute leakage is
then comparable to the physical signal, and omitting $\Pi$ inflates
$\sigma(\rho)$ by a factor of order two and makes it depend on the
finite-difference step. With Eq.~\eqref{eq:vbeta} the per-edge
uncertainties become interpretable, and they are not small: on the
full-size graphs the median $\sigma(\rho)/\rho$ runs from $0.30$ on the
email graph to $0.45$ on the Delaunay triangulation, against a few per
cent on the twelve-vertex controls.

   \section{Datasets}
   \label{sec:datasets}

We test the pipeline on one empirical network, two
geometrically-embedded networks derived from a sky catalogue, and two
small controlled graphs. This section defines each dataset and the
numerical setup; the reconstructions are reported in
Sect.~\ref{sec:results}.

\subsection{Numerical setup}
\label{sec:setup}

For each test-bed we build the true adjacency $A$, form the true
transition matrix $P_{\mathrm{true}}$ by row-normalising $A$, and use
a uniform vertex prior $p_0=\mathbf 1/n$. The co-visitation of
Eq.~\eqref{eq:covis} of length $T$ is obtained either analytically
(as $C$) or empirically by sampling $W$ finite walks and counting
observed transitions (as $\hat C$). We consider two noise regimes:
\begin{itemize}
   \item \emph{Multiplicative}:
      $\hat C_{ij}=C_{ij}\,(1+\nu\,\eta_{ij})$ with
      $\eta_{ij}\sim\mathcal N(0,1)$ i.i.d.\ and clipping to
      $\hat C_{ij}\ge 0$; $\nu$ is the noise level, kept distinct from
      the floor $\varepsilon$ of Eq.~\eqref{eq:S}.
   \item \emph{Finite walks}: $W$ trajectories of length $T$ drawn
      from $P_{\mathrm{true}}$ with starting vertices drawn from
      $p_0$; $\hat C$ is the per-walk transition-count matrix. In this
      regime the basis is restricted to the empirical support of
      $\hat C$, which for clean trajectories is a subset of the true
      edges.
\end{itemize}
The \texttt{email-Eu-core} edge list is directed. It is symmetrised once,
$A^{\mathrm{und}}_{ij}=A^{\mathrm{dir}}_{ij}\lor A^{\mathrm{dir}}_{ji}$,
before anything else. The symbol $\lor$ is the logical \emph{or} and
the operation is nothing beyond it: the undirected edge is present when
a message was sent in either direction, no counts are summed and no
weights are formed, every entry staying zero or one. Self-loops are
discarded at the same step. The walk is then sampled on
$P_{\mathrm{true}}=\mathrm{rownorm}(A^{\mathrm{und}})$, the transition
counts are accumulated as ordered pairs and symmetrised into the basis,
and $A^{\mathrm{und}}$ is also the adjacency used for scoring. The three
stages therefore share one convention, which is why the observed support
is contained in the true edge set exactly rather than approximately.
The group weights of Eq.~\eqref{eq:stiefel} are solved for whenever the
basis holds at most $700$ motifs and are set to unity above that, since
the solver stores one $m\times m$ block per vertex and the memory cost
grows as $nm^2$; all runs reported here are below the threshold and use
the solved weights.
Walk length is $T=20$ for the email subgraph and $T=16$ otherwise;
finite-walk experiments use $100$ walkers unless stated. The random
seed is fixed for reproducibility, and timings are quoted on a
laptop-class machine.

The reconstruction plots show two panels: the ground truth on the
left, and the reconstruction on the right with true positives (TP,
green), false negatives (FN, dotted grey) and false positives (FP,
dashed red). For the Delaunay and Voronoi graphs the vertex positions
are the true RA/DEC coordinates, so the agreement can be read off on
the sky, with the full galaxy point set shown in the background and
the graph vertices marked distinctly (Voronoi vertices differ from
the galaxies); for the email graph and the two controls a spring
layout of the ground truth is used and reused on the reconstruction
panel.

As a baseline we compare against graphical lasso
\citep{friedman2008}, a standard $\ell_1$-penalised
structure-learning method. It is run on the same observed
co-visitation, treated as a covariance-like matrix, and declares an
edge where the estimated precision has a non-zero off-diagonal entry;
the penalty is selected without reference to the ground truth by a
BIC-like criterion, $-\log L+k\log n$ with $k$ the number of declared
edges, over the fixed grid
$\alpha\in\{0.002,0.005,0.01,0.02,0.05,0.1,0.2\}$. The co-visitation is
symmetrised, made diagonally dominant and rescaled to unit diagonal
before it is passed to the solver, since it is not a sample covariance
and carries no sample size; the baseline is therefore a reference point
rather than a calibrated competitor. The baseline uses no plots and is recorded
alongside the fbLM result for each graph.

\subsection{Empirical network: \texttt{email-Eu-core}}
\label{sec:data-email}

The \texttt{email-Eu-core} network records email exchanges within a
European research institution, with a directed edge $(u,v)$ whenever
$u$ sent at least one message to $v$ \citep{leskovec2007}. We extract
connected subgraphs of varying size ($N=20$, $100$, and $240$) as
empirical, non-geometric test-beds with a broad degree distribution.

\subsection{COSMOS Delaunay network}
\label{sec:data-delaunay}

From a COSMOS-field galaxy catalogue \citep{smolcic2007} we take a
set of sky positions (RA, DEC) and build the Delaunay triangulation
of the point set; vertices are galaxies and edges are triangulation
links. A breadth-first subgraph of a chosen size is used for
reconstruction. Delaunay graphs are triangle-rich by construction
and probe the pipeline in the clustered regime.

\subsection{COSMOS Voronoi network}
\label{sec:data-voronoi}

From the same COSMOS point set \citep{smolcic2007} we build the
Voronoi tessellation and take its vertex-adjacency graph: vertices
are Voronoi vertices (cell corners) and edges are Voronoi ridges. Voronoi graphs are locally
sparse, dominated by chains and low-degree junctions, and share the
geometric embedding of the Delaunay case but with a very different
local structure. Boundary Voronoi vertices can extend well outside
the input point set, which is visible in the reconstruction plots.

\subsection{Controlled graphs: unicyclic and radialness}
\label{sec:data-controls}

Two $12$-vertex graphs serve as sanity checks, chosen to differ in
cycle content rather than in size. The \emph{unicyclic} control carries
$12$ edges on $12$ vertices and therefore exactly one independent
cycle: a central vertex joined to two three-leaf stars and to a
triangle, with maximum degree four. The \emph{radialness} control is a
tree, $11$ edges on $12$ vertices, a three-vertex spine carrying two,
four and three leaves with maximum degree six; having no closed path,
its co-visitation is dominated by back-and-forth walk on each edge.
Both were used throughout the development of the pipeline.

   \section{Results}
   \label{sec:results}

Table~\ref{tab:compare} summarises the reconstruction of all five
test-beds and compares the fbLM pipeline against the graphical-lasso
baseline of Sect.~\ref{sec:setup}. These runs use the finite-walk
regime with $100$ walkers, at three sizes: a small $N=20$ proxy, an
intermediate $N=100$, and a size-limited run in which the geometric
graphs are reconstructed in full (Delaunay $N=119$, Voronoi $N=223$,
both exhausting the COSMOS point set) and the empirical graph reaches
$N=240$ ($417$ edges); the controls are fixed $12$-node graphs. The
distinction matters at the smaller sizes. A subgraph of $N$ vertices is
extracted breadth-first, so the vertices on its rim keep neighbours
that were left outside while the edges to them are absent from the
extracted adjacency: their degrees are truncated, and a walk that would
have left through them stays inside instead. At full extent there is no
such rim, and the graph the walk explores is the graph being
reconstructed. At full size the
graphical-lasso reference returns MCC $0.540$ on Delaunay and $0.603$ on
the email graph, against $0.988$ and $0.967$ for fbLM; the gap is
widest on the triangle-rich graphs, although the two methods do not
receive equivalent candidate sets and the graphical-lasso result is
included only as a reference.
The reference searches all $\binom{N}{2}$ pairs and is not given the
support restriction (Sect.~\ref{sec:setup}), so the two columns are not
a like-for-like contest. fbLM holds MCC $\gtrsim 0.98$ on the geometric
graphs up to full size, and on the densest empirical graph the
residual misses are the edges the walk never reaches rather than
mis-ranked ones. The
graphical-lasso reference is faster but returns lower MCC on the larger
email and Delaunay cases.

\begin{table*}
\centering
\caption{
   Reconstruction quality across the five test-beds, fbLM versus the
   graphical-lasso baseline, in the finite-walk regime ($100$ walkers),
   at three subgraph sizes. Columns give the Matthews correlation
   coefficient (MCC) and the counts of true positives, false negatives,
   and false positives (TP/FN/FP) against the ground-truth adjacency,
   together with the fbLM fit time $t_{\mathrm{fit}}$ and the
   graphical-lasso fit time $t_{\mathrm{L}}$, in seconds on a
   laptop-class machine. The two control
   graphs (unicyclic, radialness) are fixed $12$-node graphs. The largest
   block is the size-limited run: the Delaunay ($N=119$) and Voronoi
   ($N=223$) test-beds exhaust the COSMOS point set and are reconstructed
   in full, with zero subgraph boundary. fbLM holds MCC $\gtrsim 0.98$
   on the geometric graphs up to full size and degrades gracefully to
   the densest empirical graph ($N=240$, $417$ edges), where the residual
   misses are the edges the walk never traverses, while the
   graphical-lasso baseline
   reference returns substantially lower MCC on the triangle-rich
   empirical and Delaunay graphs ($0.603$ and $0.540$). For the smallest size-class ($N=20$, and the
   fixed $12$-node controls) every MCC and time is reported as the
   median over $N_{\mathrm{sim}}=100$ independently resampled datasets,
   with asymmetric $16$th/$84$th-percentile bars
   $x^{+(p_{84}-\mathrm{med})}_{-(\mathrm{med}-p_{16})}$; the fit-time
   spread reflects data-resampling variability, not machine jitter
   alone. For the larger sizes ($N=100$ and the full/size-limited run)
   the repeated simulation is impractical (a single fbLM fit already
   costs up to $\sim7$ minutes at $N=240$), so those rows quote a
   single representative run without uncertainties. The fbLM fit time
   grows with the number of motifs.
   The column $m$ is the size of the candidate basis, and in the
   finite-walk regime it also defines a trivial baseline: declaring an
   edge wherever the walk traversed one gives $\mathrm{TP}=m$,
   $\mathrm{FN}=E-m$ and $\mathrm{FP}=0$ by construction, since for clean
   trajectories the observed support is contained in the true edge set.
   fbLM reproduces that baseline exactly on nine of the eleven rows and
   falls two edges short on \texttt{email-Eu-core} at $N=20$ and one at
   $N=100$, where the support baseline would score $1.000$ against
   $0.953$ and $0.997$. In this regime, therefore, the binary adjacency
   is set by coverage and not by the fit; what the fit adds is the
   weighting of the supported pairs and the per-edge uncertainty of
   Sect.~\ref{sec:res-unc}, and the estimator is exercised on a
   basis containing genuine non-edges only in the full-basis experiments,
   the coverage study of Fig.~\ref{fig:cost} and the multiplicative-noise
   regime of Table~\ref{tab:mult}; Sect.~\ref{sec:res-mult} sets out what
   the latter does and does not establish. The
   graphical-lasso baseline receives no support restriction and searches
   all $\binom{N}{2}$ pairs, so its column is not a like-for-like
   comparison and is reported for reference rather than as a contest.
}
\label{tab:compare}
{\setlength{\tabcolsep}{5pt}
\begin{tabular}{llccccccccc}
\toprule
 & & & & & \multicolumn{3}{c}{fbLM} & \multicolumn{3}{c}{graphical lasso} \\
\cmidrule(lr){6-8}\cmidrule(lr){9-11}
Graph & $N$ & $E$ & $m$ & & MCC & TP/FN/FP & $t_{\mathrm{fit}}$\,[s] & MCC & TP/FN/FP & $t_{\mathrm{L}}$\,[s] \\
\midrule
\multicolumn{11}{l}{\emph{$N=20$}}\\
\texttt{email-Eu-core} & 20 & 25 & 25 & & $0.953^{+0.000}_{-0.000}$ & 23/2/0    & $0.409^{+0.002}_{-0.010}$ & $0.831^{+0.025}_{-0.026}$ & 19/6/0    & $0.0130^{+0.0005}_{-0.0001}$ \\[2pt]
Delaunay (COSMOS)      & 20 & 45 & 45 & & $1.000^{+0.000}_{-0.000}$ & 45/0/0    & $0.948^{+0.021}_{-0.013}$ & $0.683^{+0.032}_{-0.033}$ & 21/24/0   & $0.0141^{+0.0008}_{-0.0005}$ \\[2pt]
Voronoi (COSMOS)       & 20 & 24 & 24 & & $1.000^{+0.000}_{-0.000}$ & 24/0/0    & $0.400^{+0.004}_{-0.002}$ & $1.000^{+0.000}_{-0.000}$ & 24/0/0    & $0.0195^{+0.0012}_{-0.0015}$ \\
\midrule
\multicolumn{11}{l}{\emph{$N=100$}}\\
\texttt{email-Eu-core} & 100 & 173 & 173 & & 0.997 & 172/1/0   & 36.1  & 0.620 & 68/105/0  & 0.15 \\
Delaunay (COSMOS)      & 100 & 269 & 266 & & 0.994 & 266/3/0   & 83.3  & 0.577 & 93/176/0  & 0.09 \\
Voronoi (COSMOS)       & 100 & 138 & 138 & & 1.000 & 138/0/0   & 25.8  & 0.981 & 133/5/0   & 0.17 \\
\midrule
\multicolumn{11}{l}{\emph{full / size-limited}}\\
\texttt{email-Eu-core} & 240 & 417 & 390 & & 0.967 & 390/27/0  & 422.8 & 0.603 & 153/264/0 & 1.10 \\
Delaunay (COSMOS)      & 119 & 341 & 333 & & 0.988 & 333/8/0   & 164.7 & 0.540 & 103/238/0 & 0.19 \\
Voronoi (COSMOS)       & 223 & 328 & 313 & & 0.977 & 313/15/0  & 262.4 & 0.910 & 272/56/0  & 1.36 \\
\midrule
\multicolumn{11}{l}{\emph{controls (fixed $12$-node)}}\\
unicyclic (control)    & 12 & 12 & 12 & & $1.000^{+0.000}_{-0.000}$ & 12/0/0    & $0.183^{+0.001}_{-0.000}$ & $1.000^{+0.000}_{-0.000}$ & 12/0/0    & $0.0141^{+0.0020}_{-0.0016}$ \\[2pt]
radialness (control)   & 12 & 11 & 11 & & $1.000^{+0.000}_{-0.000}$ & 11/0/0    & $0.173^{+0.001}_{-0.001}$ & $1.000^{+0.000}_{-0.055}$ & 10/1/0    & $0.0722^{+0.0233}_{-0.0453}$ \\
\bottomrule
\end{tabular}}
\end{table*}

\subsection{Empirical network}
\label{sec:res-email}

\begin{figure*}
\centering
\includegraphics[width=\textwidth]{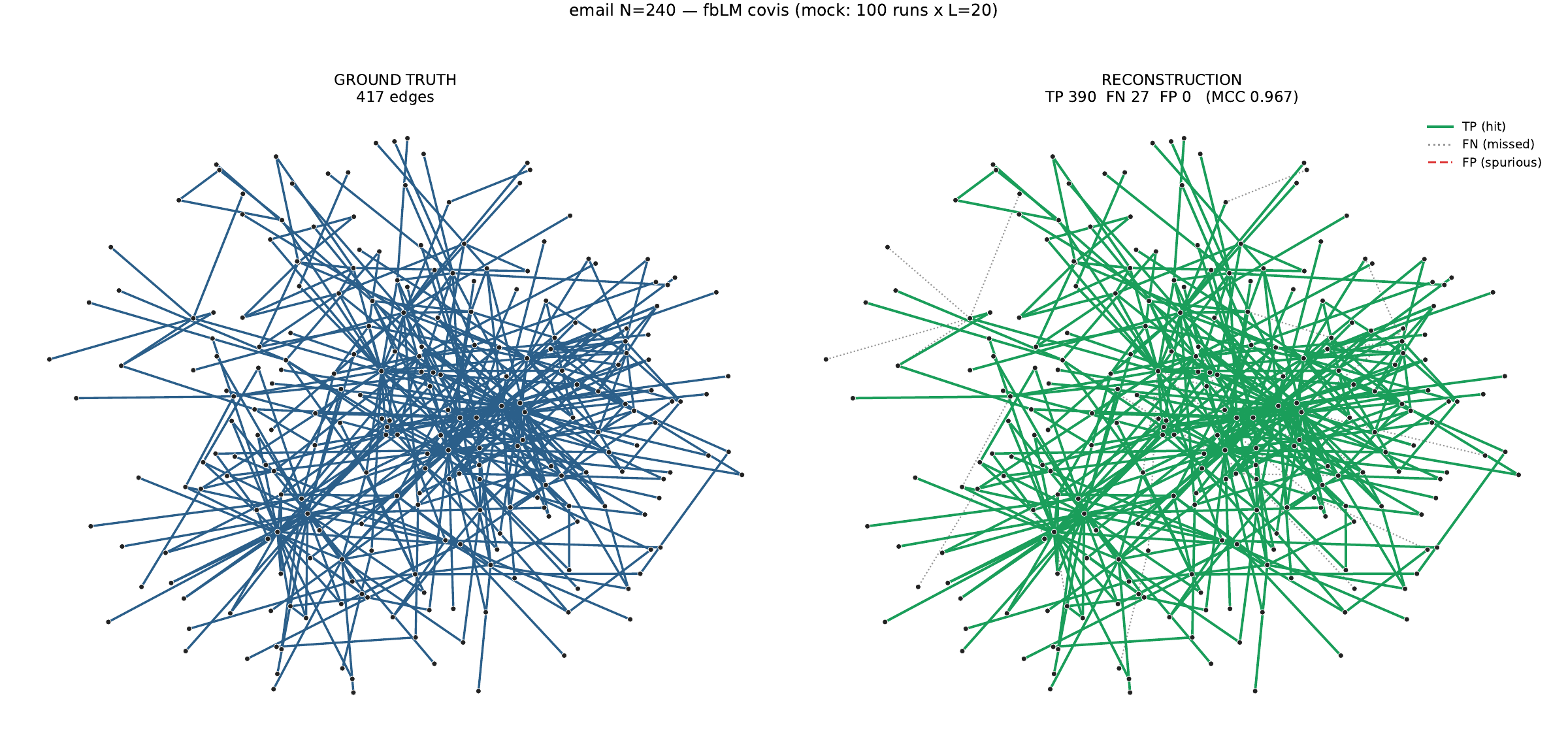}
\caption{
   Reconstruction of the \texttt{email-Eu-core} subgraph. Left:
   ground-truth adjacency. Right: reconstruction with TP (green), FN
   (grey dotted), and FP (red dashed); the layout is a spring layout
   of the ground truth reused on both panels. Aggregate statistics
   (TP/FN/FP and MCC) are printed in the panel title.
}
\label{fig:email}
\end{figure*}

Figure~\ref{fig:email} shows the reconstruction of the largest
\texttt{email-Eu-core} subgraph ($N=240$, $417$ edges). The residual
misses are set by coverage rather than by the estimator: with $100$
walks of length $20$ the trajectories never reach $10$ of the $240$
vertices and never traverse $27$ of the $417$ edges, and those edges
are exactly the false negatives. No false positives occur, and none can:
for clean trajectories the walk moves only along true edges, so the
observed support is contained in the true edge set and the candidate
basis inherits that containment.

\subsection{Geometric networks}
\label{sec:res-cosmos}

\begin{figure*}
\centering
\includegraphics[width=\textwidth]{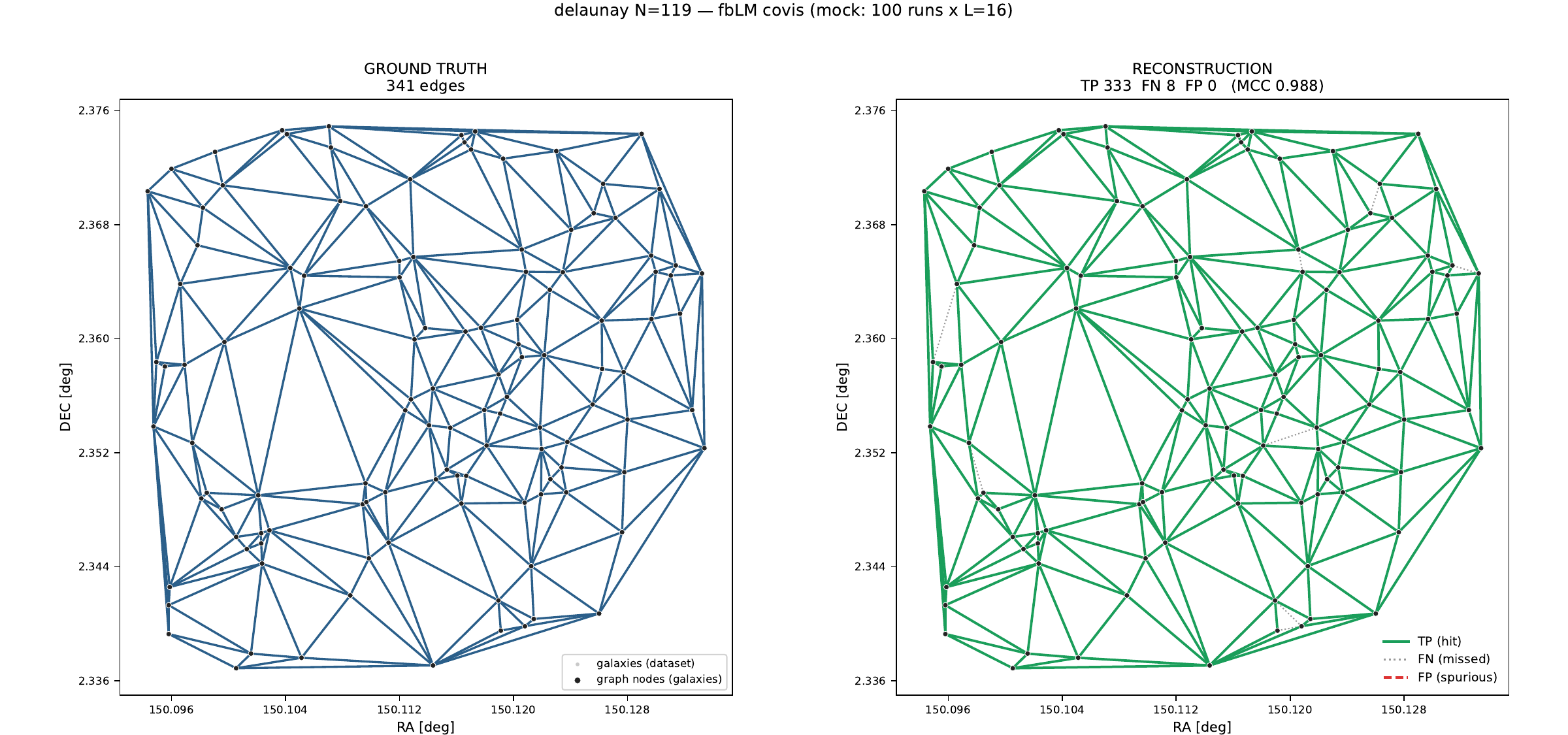}
\caption{
   Reconstruction of the COSMOS Delaunay subgraph on true RA/DEC
   coordinates. Left: ground truth. Right: reconstruction with TP
   (green), FN (grey dotted), FP (red dashed).
}
\label{fig:delaunay}
\end{figure*}

\begin{figure*}
\centering
\includegraphics[width=\textwidth]{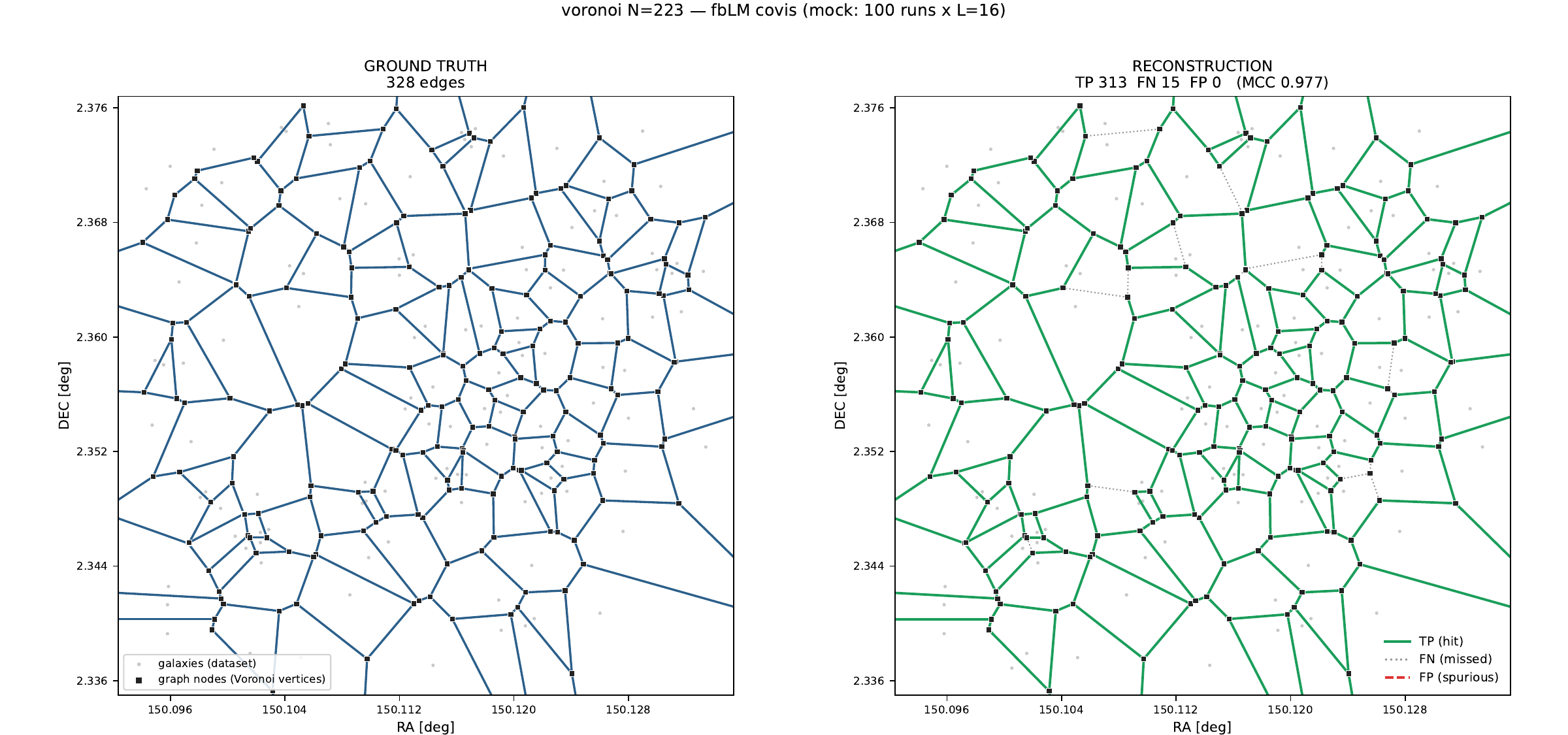}
\caption{
   Reconstruction of the COSMOS Voronoi subgraph on true RA/DEC
   coordinates. Left: ground truth. Right: reconstruction with TP
   (green), FN (grey dotted), FP (red dashed). Boundary Voronoi
   vertices can extend well outside the input point set.
}
\label{fig:voronoi_recon}
\end{figure*}

Delaunay and Voronoi reconstructions are shown on true RA/DEC
coordinates in Figs.~\ref{fig:delaunay} and \ref{fig:voronoi_recon}.
In the size-limited run both are reconstructed at full extent, with no
subgraph boundary (Delaunay $N=119$, MCC $=0.988$; Voronoi $N=223$,
MCC $=0.977$; see Table~\ref{tab:compare}); the triangulated Delaunay
structure,
the structure on which additive node-potential models fail, is
recovered by fbLM without systematic error, the graphical-lasso
reference returning MCC $=0.540$ at this size, and
the sparse Voronoi structure is tracked closely on the sky.

\subsection{Controlled graphs}
\label{sec:res-controls}

\begin{figure*}
\centering
\includegraphics[width=\textwidth]{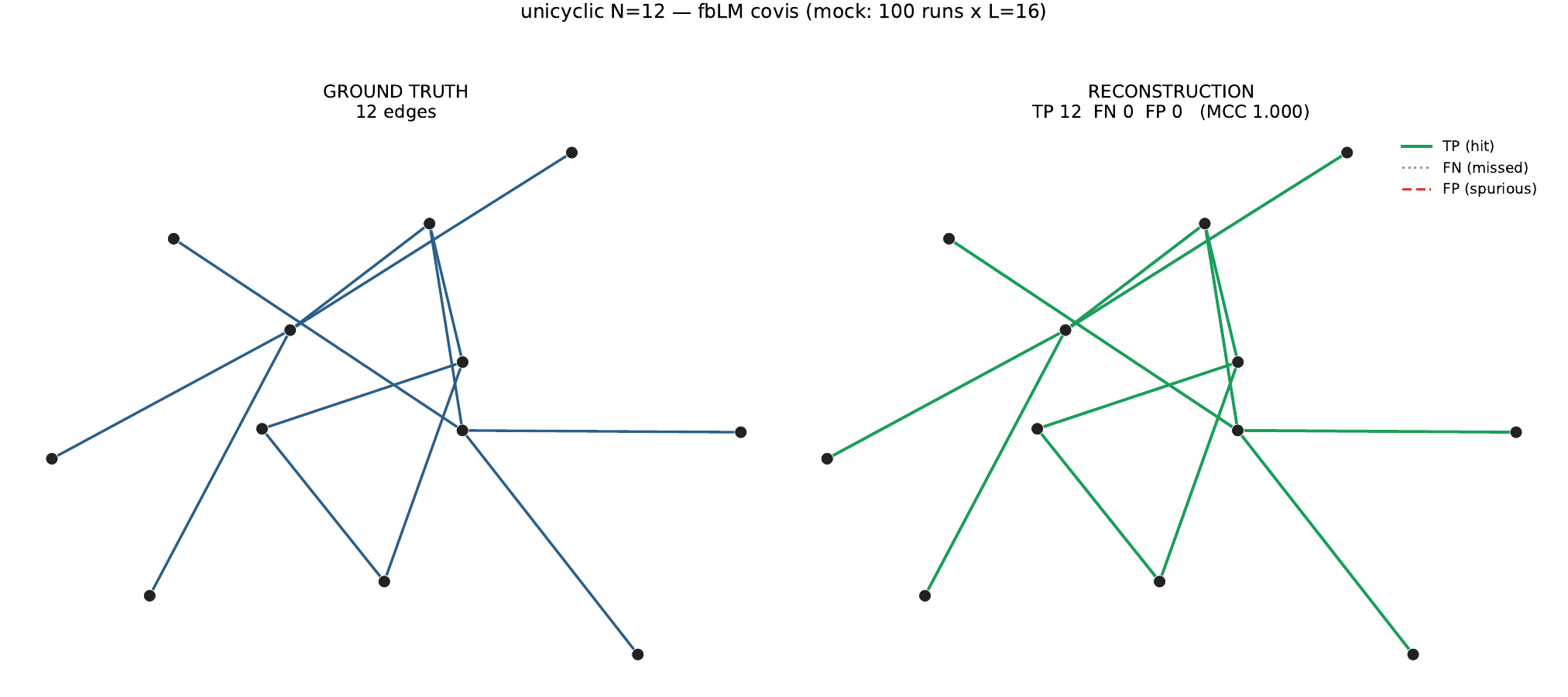}
\caption{
   Reconstruction of the $12$-vertex unicyclic control graph. Left: ground
   truth. Right: reconstruction with TP (green), FN (grey dotted), FP
   (red dashed).
}
\label{fig:tree}
\end{figure*}

\begin{figure*}
\centering
\includegraphics[width=\textwidth]{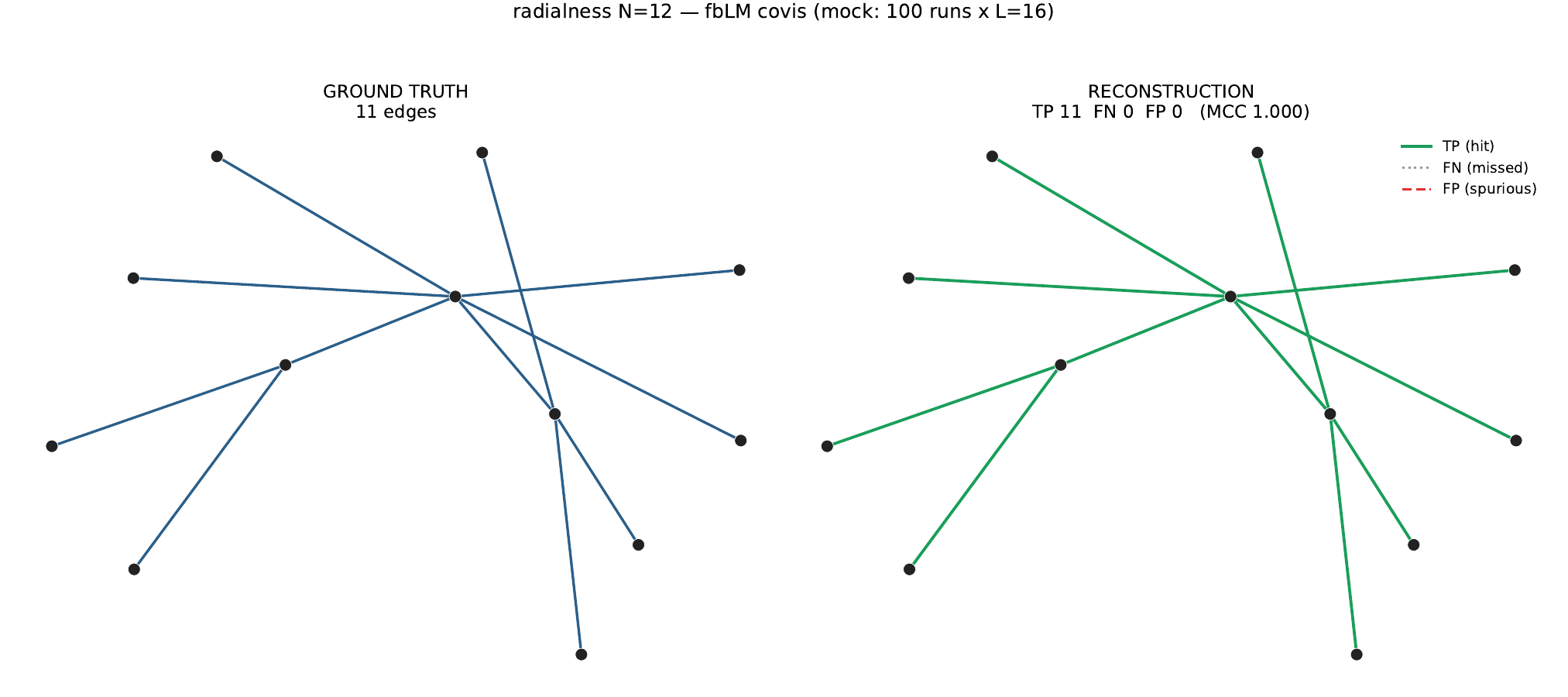}
\caption{
   Reconstruction of the $12$-vertex radialness control graph, a tree.
   Left:
   ground truth. Right: reconstruction with TP (green), FN (grey
   dotted), FP (red dashed).
}
\label{fig:radialness}
\end{figure*}

The unicyclic and radialness controls (Figs.~\ref{fig:tree} and
\ref{fig:radialness}) are reconstructed exactly, as expected for such
simple structures, confirming that the readout separates edges
cleanly in the low-degree regime.

\subsection{Coverage-limited cost curve}
\label{sec:res-cost}

\begin{figure}
\centering
\includegraphics[width=0.95\linewidth]{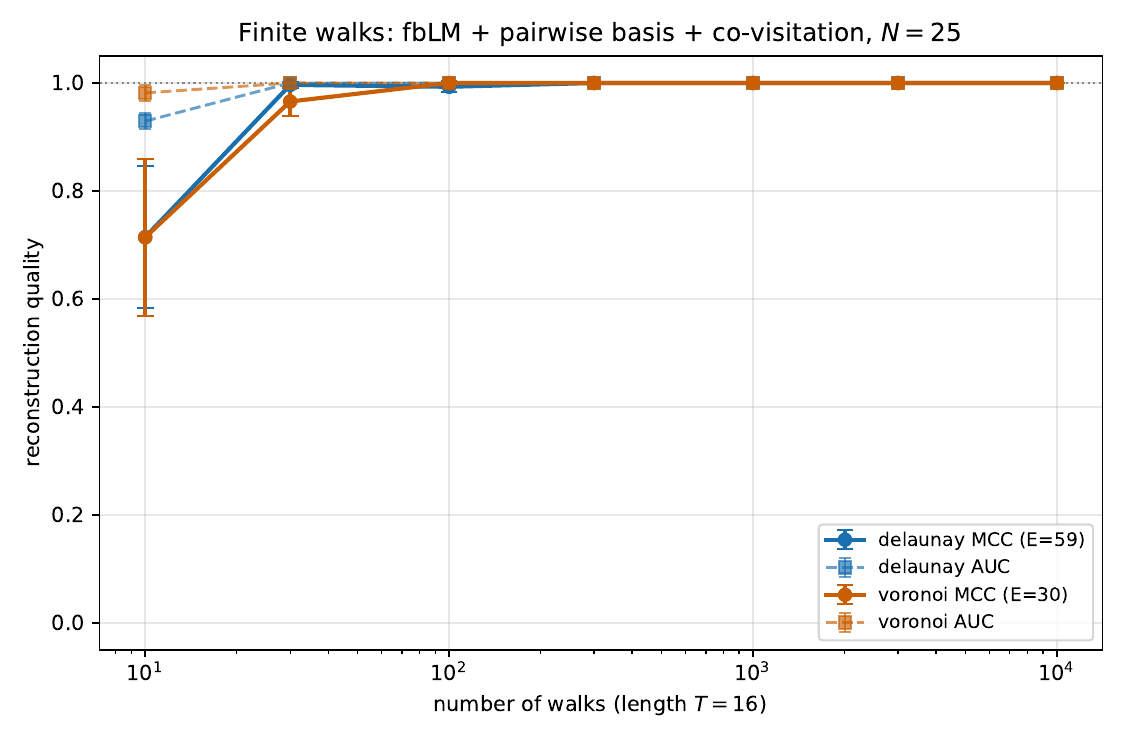}
\caption{
   Reconstruction MCC versus number of walks at fixed length $T=16$.
   Solid curves with circles show MCC, dashed curves with squares show
   AUC; symbols are means over three independent samplings and error
   bars span one standard deviation. Near-perfect reconstruction is
   approached after $30$ walks for both Delaunay and Voronoi at $N=25$,
   and the curves hold at unity out to $10^{4}$ walks. At $10$ walks the
   AUC is already above $0.92$ while the MCC is near $0.71$: the edge
   ranking survives a coverage at which the threshold does not.
}
\label{fig:cost}
\end{figure}

For the finite-walk regime, Fig.~\ref{fig:cost} reports reconstruction
quality against the number of walks at fixed length $T=16$, over three
independent samplings per point. This experiment fits the full pairwise
basis rather than the observed support, so every pair carries a
parameter and the AUC below is a genuine ranking over all
$\binom{n}{2}$ pairs; it is the one place in the paper where the two
basis conventions of Sect.~\ref{sec:forward} differ. Near-perfect reconstruction is reached
already at $30$ walks, some $480$ transitions, for both test-beds at
$N=25$ (MCC $0.996\pm0.005$ for Delaunay and $0.966\pm0.028$ for
Voronoi); the curves settle at $1.000$ by a few hundred walks and stay
there out to $10^{4}$, so nothing is lost by oversampling. At $10$ walks
the threshold has become unreliable (MCC $0.714$ for both) while the
edge \emph{ranking} still holds, AUC $0.929\pm0.014$ for Delaunay and
$0.982\pm0.014$ for Voronoi, which is why both quantities are shown. The
gap between them should be read with care: an edge the walk never
traverses has a target of zero, like a non-edge, so the AUC at low
coverage is close to what coverage alone would give if unseen edges were
placed at chance among the non-edges. Across seeds it lands within a few
per cent of that value in either direction, so the ordering does carry
information beyond coverage, but the margin is small, and what the gap
mainly shows is that the threshold fails before the ordering does.

The limiting factor is walk coverage rather than the estimator, and the
reconstructions of Sect.~\ref{sec:res-email} make that quantitative.
Every false negative in Table~\ref{tab:compare} is, to within one or two
edges, an edge the walk never traversed: $27$ untraversed against $27$
missed for the $N=240$ email graph, $8$ against $8$ for Delaunay at
$N=119$, $15$ against $15$ for Voronoi at $N=223$. The only exceptions
are one edge at $N=100$ and two at $N=20$ on the email graph, which were
traversed but fell below the readout threshold. What the walk does not
visit, no fit recovers; what it visits, the fit almost always keeps.

\subsection{Multiplicative-noise regime}
\label{sec:res-mult}

The results above all come from sampled trajectories. The second
regime defined in Sect.~\ref{sec:setup} perturbs the analytic
co-visitation directly, $\hat C = C\,(1+\nu\eta)$ with
$\eta\sim\mathcal N(0,1)$ and clipping at zero, which isolates the
fitter's tolerance to noise on the observable from the question of
coverage: here every pair is present in the basis, so nothing is
unreachable and any failure is the estimator's own. Table~\ref{tab:mult}
reports it. Both geometric test-beds are recovered exactly at every seed and at
every size, with no misclassified edge in any run. It is worth being
explicit about why that result is easier than it looks, because the
uniformity of the table invites the opposite reading.

The analytic co-visitation factorises as
$C_{ij}=P_{ij}\sum_t p_t(i)$ with $P=\mathrm{rownorm}(A)$, so its support
is exactly the true edge set, and a multiplicative perturbation cannot
change that: $C_{ij}=0$ forces $\hat C_{ij}=0$, and clipping at zero
would require a deviate below $-1/\nu$, which at these noise
levels never occurs. The fitting target therefore carries exact zeros at
precisely the non-edges, and the pattern of zeros already encodes the
answer. What the fit has to do is not separate weak edges from strong
non-edges, since no such cases exist, but avoid destroying a separation
it is handed. It does so with room to spare: at $N=25$ the fitted
weights of the $241$ non-edges all settle within two per cent of one
another, a factor of eighteen below the weakest edge, and in the readout
the largest non-edge coupling sits six times below the threshold.

The regime therefore measures tolerance to amplitude noise on the
non-zero entries rather than the ability to discriminate edges from
non-edges. The lower block of Table~\ref{tab:mult} locates that
tolerance by holding $N=25$ and raising the noise to $40\%$, four times
the level used above, and the outcome is instructive in a way the single
point was not. The AUC stays at $1.000$ at every level, and no false
positive occurs anywhere; what degrades is the Matthews coefficient, from
$1.000$ to $0.939$ on the Delaunay graph, entirely through edges that
drop below the readout threshold while still ranking above every
non-edge. The failure mode under amplitude noise is thus the same one the
coverage study shows under sparse sampling: the threshold gives way
before the ordering does.

Two caveats belong with that block. Three realisations per level are too
few to resolve a curve, and the scatter shows it: the Voronoi graph loses
one edge at $20\%$ and none at $30\%$ or $40\%$, which is not a noise
effect but a converged fit at a poor local minimum. Its final $\chi^{2}$
is fifteen times worse than at $30\%$ and does not improve when the
iteration budget is tripled, so at these amplitudes the optimiser's
choice of minimum varies as much as the perturbation does. Separating the
two would need many more realisations, or restarts, and we have not done
it here. The upper size, in turn, is set by the solver and not by the
method: the group-weight step holds one $m\times m$ block per vertex, so
its memory grows as $nm^{2}$, and $N=37$ is the largest full-pairwise
basis for which the weights are still solved for rather than set to
unity. The contrast with the
finite-walk columns of Table~\ref{tab:compare} is the point: once
coverage is removed as a constraint, the reconstruction is exact, which
locates the residual errors of the sampled regime in the sampling
rather than in the fit.

\begin{table}
\centering
\caption{
   Reconstruction of the COSMOS Delaunay and Voronoi test-beds in the
   multiplicative-noise regime, where the analytic co-visitation is
   perturbed entry-wise and the motif basis spans all $\binom{N}{2}$
   pairs, so that $m$ is the number of fitted parameters and $m-E$ the
   number of non-edge parameters the fit has to suppress. The upper block
   fixes the noise and varies the size, the lower fixes $N=25$ and varies
   the noise. TP/FN/FP is quoted for the worst of the three realisations;
   the false-positive count is zero in every run at every noise level, so
   all degradation is in the false negatives. The AUC over all
   $\binom{N}{2}$ pairs is $1.000\pm0.000$ throughout, including the runs
   where the MCC falls. Values are mean $\pm$ standard deviation over three
   independent noise realisations; TP/FN/FP are identical in every
   realisation and are quoted once.
}
\label{tab:mult}
{\setlength{\tabcolsep}{3pt}
\begin{tabular}{lcccccc}
\toprule
Graph & $N$ & $E$ & $m$ & noise & MCC & TP/FN/FP \\
\midrule
\multicolumn{7}{l}{\emph{size at fixed noise}}\\
Delaunay & 12 & 22 &  66 & $10\%$ & $1.000\pm0.000$ & 22/0/0 \\
Voronoi  & 12 & 13 &  66 & $10\%$ & $1.000\pm0.000$ & 13/0/0 \\
Delaunay & 25 & 59 & 300 & $12\%$ & $1.000\pm0.000$ & 59/0/0 \\
Voronoi  & 25 & 30 & 300 & $12\%$ & $1.000\pm0.000$ & 30/0/0 \\
Delaunay & 37 & 92 & 666 & $12\%$ & $1.000\pm0.000$ & 92/0/0 \\
Voronoi  & 37 & 47 & 666 & $12\%$ & $1.000\pm0.000$ & 47/0/0 \\
\midrule
\multicolumn{7}{l}{\emph{noise at $N=25$}}\\
Delaunay & 25 & 59 & 300 & $10\%$ & $1.000\pm0.000$ & 59/0/0 \\
Voronoi  & 25 & 30 & 300 & $10\%$ & $1.000\pm0.000$ & 30/0/0 \\
Delaunay & 25 & 59 & 300 & $20\%$ & $1.000\pm0.000$ & 59/0/0 \\
Voronoi  & 25 & 30 & 300 & $20\%$ & $0.994\pm0.009$ & 29/1/0 \\
Delaunay & 25 & 59 & 300 & $30\%$ & $0.982\pm0.013$ & 56/3/0 \\
Voronoi  & 25 & 30 & 300 & $30\%$ & $1.000\pm0.000$ & 30/0/0 \\
Delaunay & 25 & 59 & 300 & $40\%$ & $0.939\pm0.045$ & 49/10/0 \\
Voronoi  & 25 & 30 & 300 & $40\%$ & $1.000\pm0.000$ & 30/0/0 \\
\bottomrule
\end{tabular}}
\end{table}

\subsection{Fit quality}
\label{sec:res-fit}

\begin{figure}
\centering
\includegraphics[width=0.95\linewidth]{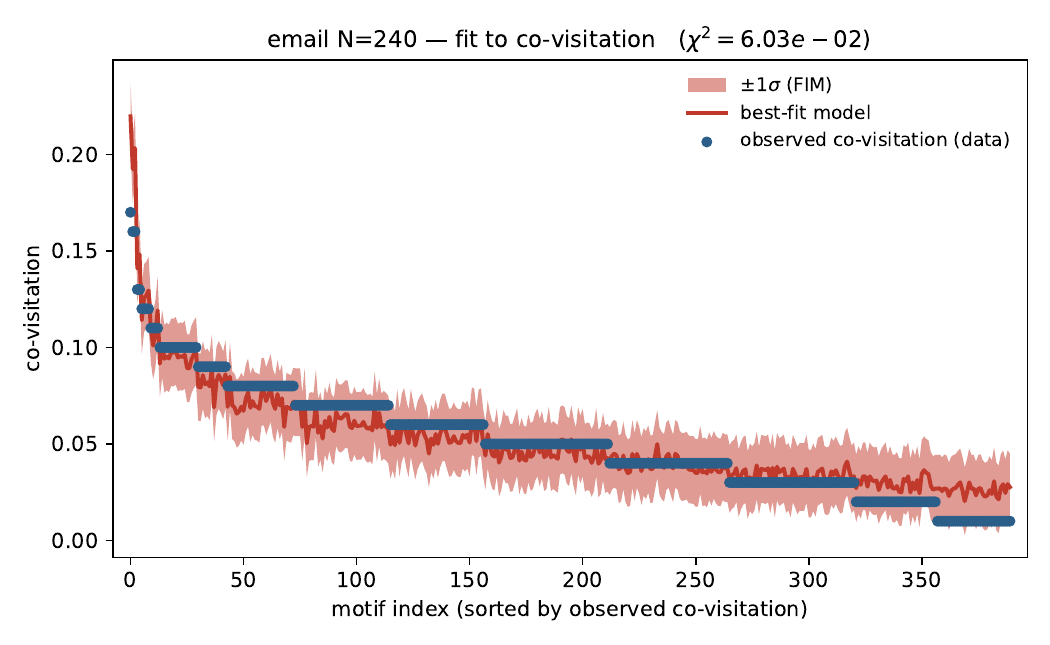}
\caption{Fit to the co-visitation for the \texttt{email-Eu-core}
   subgraph: observed (points) vs.\ best-fit model (curve).}
\label{fig:fit-email}
\end{figure}

\begin{figure}
\centering
\includegraphics[width=0.95\linewidth]{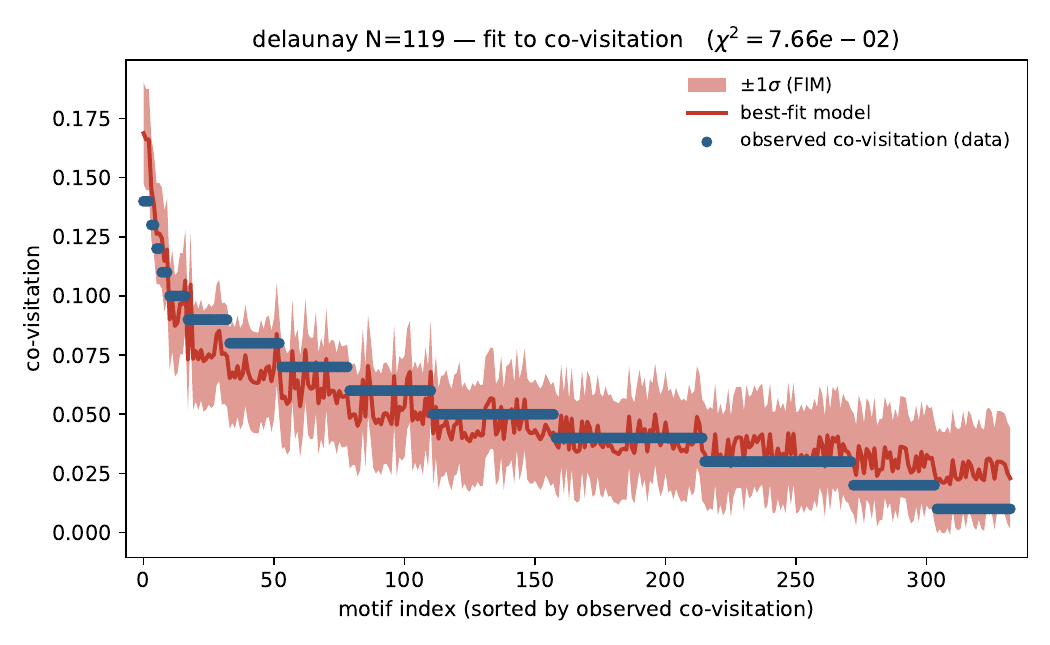}
\caption{Fit to the co-visitation for the COSMOS Delaunay subgraph.}
\label{fig:fit-delaunay}
\end{figure}

\begin{figure}
\centering
\includegraphics[width=0.95\linewidth]{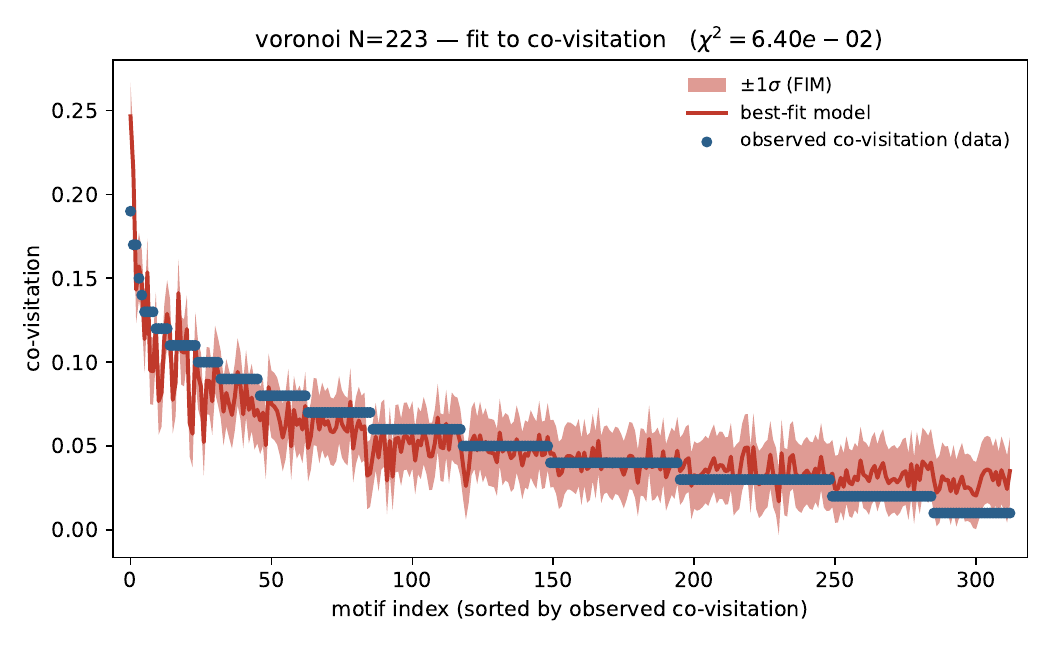}
\caption{Fit to the co-visitation for the COSMOS Voronoi subgraph.}
\label{fig:fit-voronoi}
\end{figure}

\begin{figure}
\centering
\includegraphics[width=0.95\linewidth]{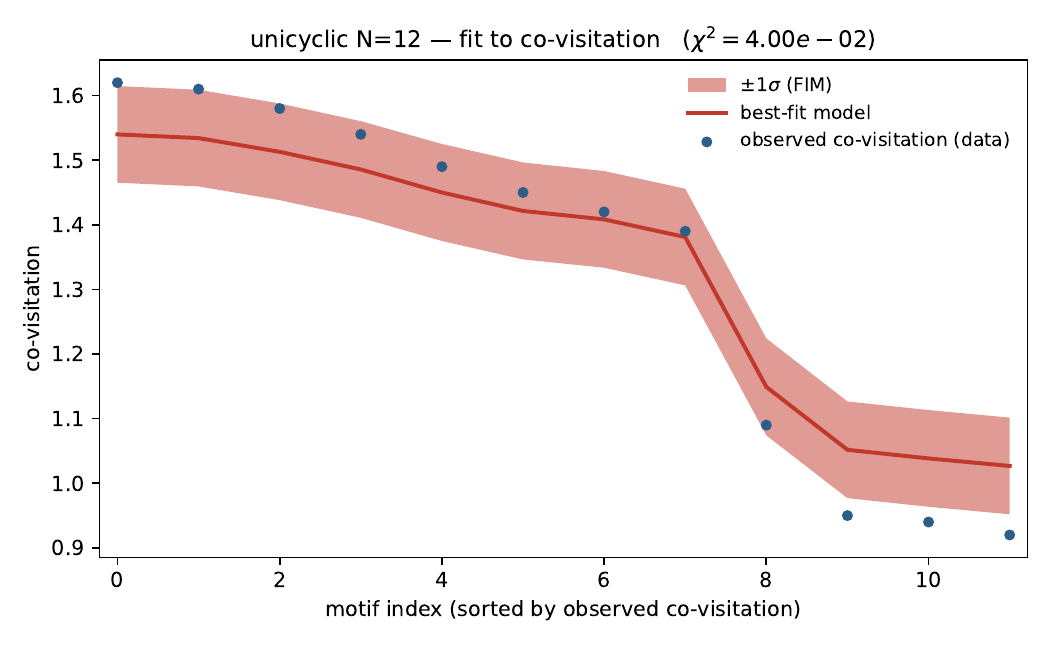}
\caption{Fit to the co-visitation for the unicyclic control graph.}
\label{fig:fit-tree}
\end{figure}

\begin{figure}
\centering
\includegraphics[width=0.95\linewidth]{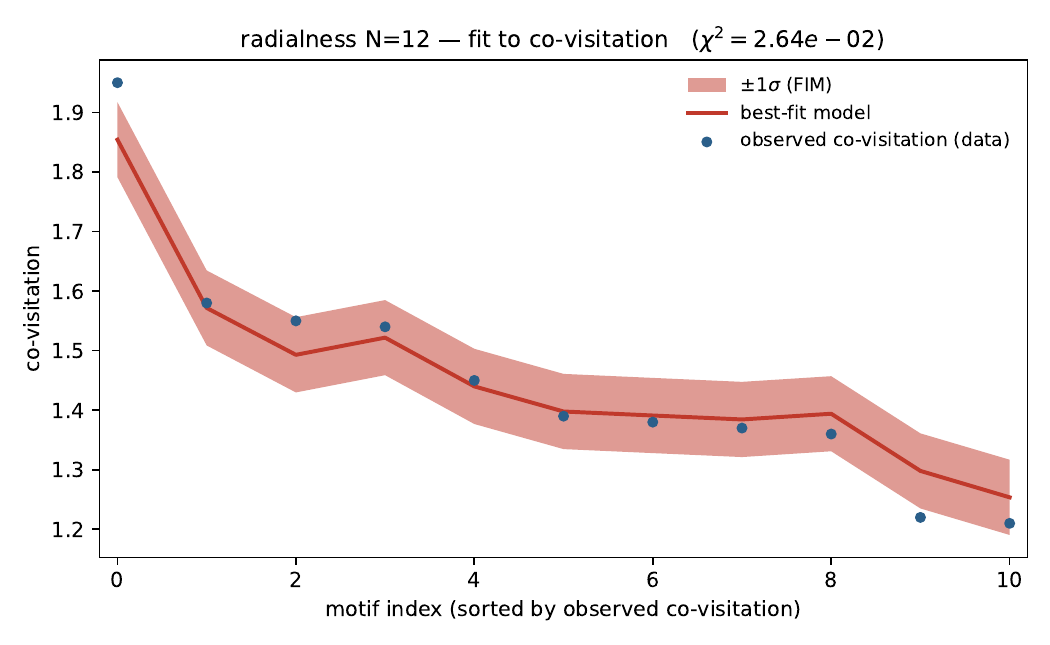}
\caption{Fit to the co-visitation for the radialness control graph.}
\label{fig:fit-radialness}
\end{figure}

Figures~\ref{fig:fit-email}--\ref{fig:fit-radialness} show, for each
test-bed, how well the fitted model reproduces the data it was given:
the observed co-visitation from the runs (points) against the
best-fit model co-visitation $C(\beta_{\mathrm{fit}})$ (curve), taken
over the motif pairs and sorted by the observed value. The shaded band
is the $\pm1\sigma$ prediction uncertainty propagated from the Fisher
information: the parameter covariance $V_\beta$ of Eq.~\eqref{eq:vbeta}
is pushed through the per-pair Jacobians to the covariance of the
predictions, whose diagonal gives $\sigma$. The best-fit curve tracks the
observed co-visitation across the whole range of pair strengths, inside
the band over most of it, the few strongest pairs being over-predicted;
the final $\chi^2$ is quoted in each panel. The observed values are
counts divided by the number of walks and therefore lie on the grid
$k/W$, which is why they appear as horizontal steps rather than as a
smooth curve.

\subsection{Per-edge uncertainty}
\label{sec:res-unc}

\begin{figure}
\centering
\includegraphics[width=0.95\linewidth]{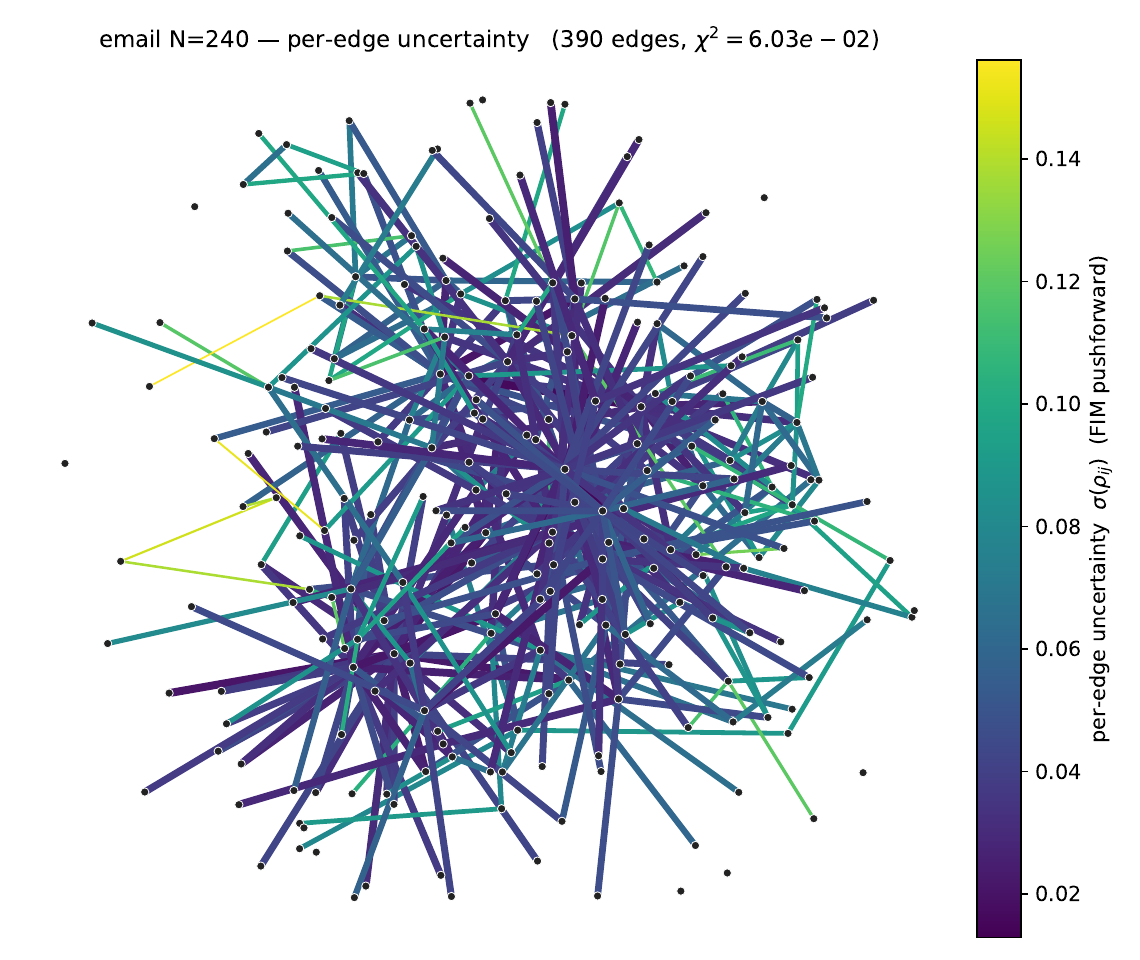}
\caption{Per-edge uncertainty $\sigma(\rho_{ij})$ for the
   \texttt{email-Eu-core} subgraph: reconstructed edges coloured by the
   FIM-propagated readout uncertainty (brighter/thinner = less certain).
   $\sigma(\rho_{ij})$ spans $0.013$ to $0.156$ against couplings
   $\rho_{ij}$ of $0.027$ to $0.778$, both in the same units.}
\label{fig:unc-email}
\end{figure}

\begin{figure}
\centering
\includegraphics[width=0.95\linewidth]{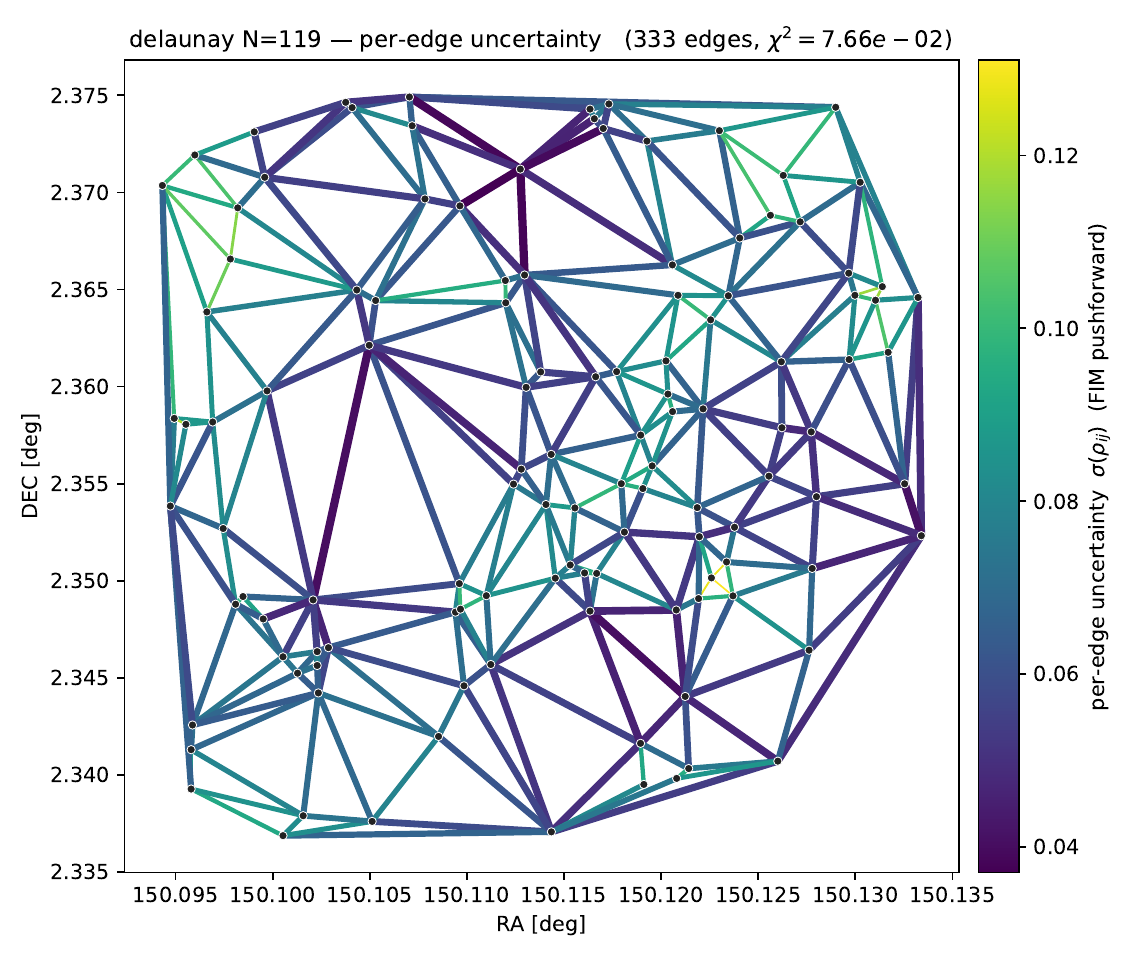}
\caption{Per-edge uncertainty for the COSMOS Delaunay subgraph
   (RA/DEC layout, dataset galaxies in the background).
   $\sigma(\rho_{ij})$ spans $0.037$ to $0.131$ against couplings
   $\rho_{ij}$ of $0.068$ to $0.506$, both in the same units.}
\label{fig:unc-delaunay}
\end{figure}

\begin{figure}
\centering
\includegraphics[width=0.95\linewidth]{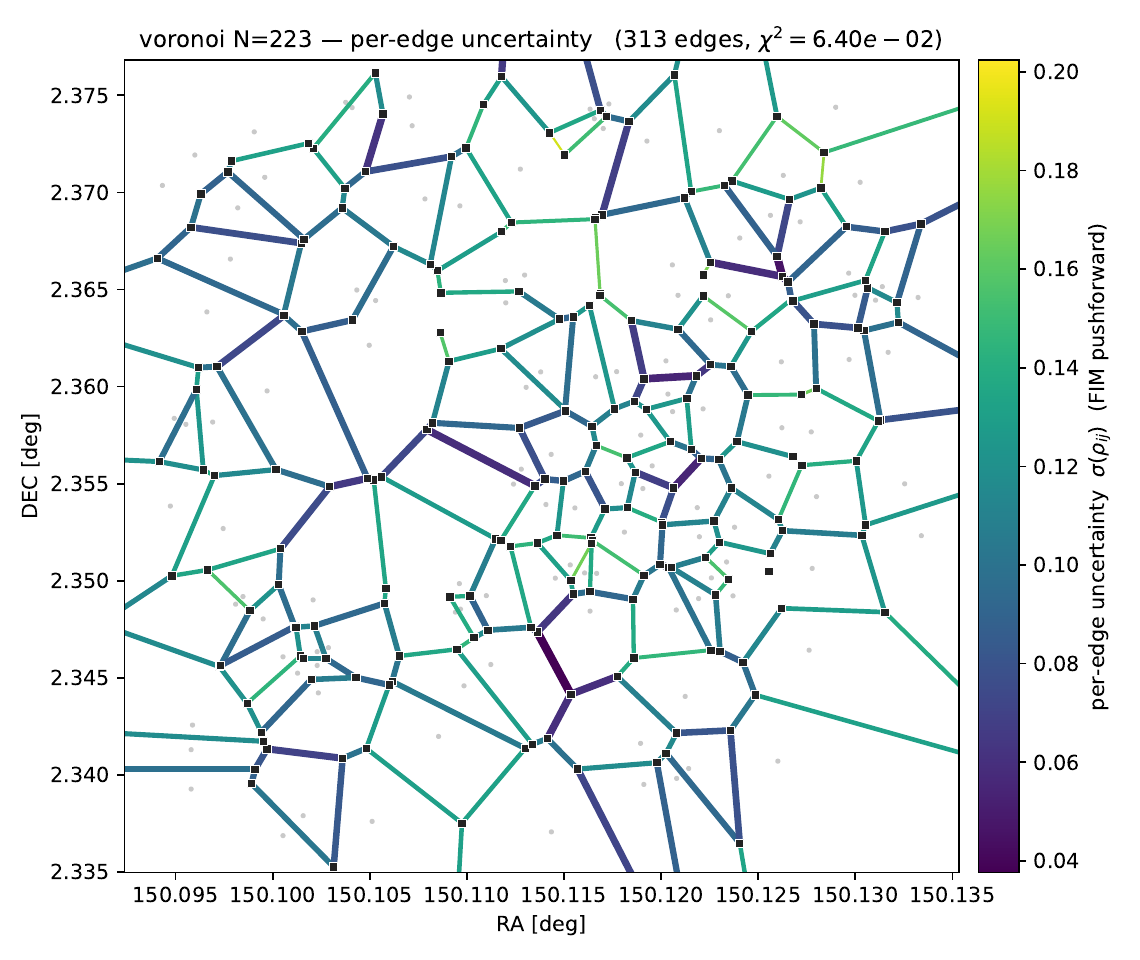}
\caption{Per-edge uncertainty for the COSMOS Voronoi subgraph
   (RA/DEC layout). $\sigma(\rho_{ij})$ spans $0.038$ to $0.202$ against
   couplings $\rho_{ij}$ of $0.110$ to $0.790$, both in the same units.}
\label{fig:unc-voronoi}
\end{figure}

The fit band of Sec.~\ref{sec:res-fit} quantifies uncertainty on the
\emph{observable} (the co-visitation). To attach an uncertainty to each
\emph{reconstructed edge}, the same parameter covariance $V_\beta$ of
Eq.~\eqref{eq:vbeta} is pushed forward not to
the co-visitation but through the per-node readout
$\rho_{ij}=W_{ij}/\sqrt{s_i s_j}$ itself:
\begin{equation}
   \Sigma_{\rho}
   = \mathbf J_{\rho}\, V_\beta\, \mathbf J_{\rho}^{\top},
   \qquad
   (\mathbf J_{\rho})_{e,k}
   = \frac{\partial \rho_{e}}{\partial \beta_k},
\end{equation}
so that $\sigma(\rho_{ij})=\sqrt{[\Sigma_\rho]_{ee}}$ is a one-standard-deviation
uncertainty for the candidate edge $e=(i,j)$. It is an uncertainty on
the \emph{coupling}, not on the binary decision: the classification
depends on the margin $\rho_{ij}-\tfrac12(\bar\rho_i+\bar\rho_j)$, and
while $\mathbf J_\rho$ carries the dependence of the strengths $s_i$ on
all parameters, the threshold $\bar\rho$ is not propagated. Because the readout is the
only place the true adjacency could enter and it does not, this is an
$A$-free per-edge coupling uncertainty: it uses only the fitted model's own Fisher
geometry. Figures~\ref{fig:unc-email}--\ref{fig:unc-voronoi} draw the
reconstructed graph on the same layout as the reconstruction panels, with
each edge coloured by $\sigma(\rho_{ij})$ (and drawn thicker when more
certain). The uncertainty concentrates where it should: on the
low-co-visitation, boundary-touching edges, while the well-traversed
interior edges are pinned down tightly. The colour scale is in the units
of $\rho$ itself, so it should be read against the range of $\rho$ quoted
in each caption rather than in absolute terms. At the sizes shown the
relative uncertainty is substantial: the median
$\sigma(\rho_{ij})/\rho_{ij}$ is $0.30$ for the email graph, $0.35$ for
Voronoi and $0.45$ for the triangle-rich Delaunay graph, and the least
determined single coupling, on the Voronoi graph, reaches
$\sigma(\rho)=\rho$; on the twelve-vertex controls it falls to a few per
cent.

That an edge set recovered at MCC $0.988$ should rest on couplings
individually uncertain at the $45\%$ level is not a contradiction but a
consequence of how the readout is built. The decision compares
$\rho_{ij}$ against the mean coupling at its own endpoints, and the
dominant modes of $V_\beta$ move whole neighbourhoods of couplings
together, so the common part cancels in the comparison while surviving
in $\sigma(\rho)$. It is a further reason to read $\sigma(\rho)$ as an
uncertainty on the coupling and not on the classification.

\subsection{Motif bases}
\label{sec:res-motifs}

\begin{figure*}
\centering
\includegraphics[width=\textwidth]{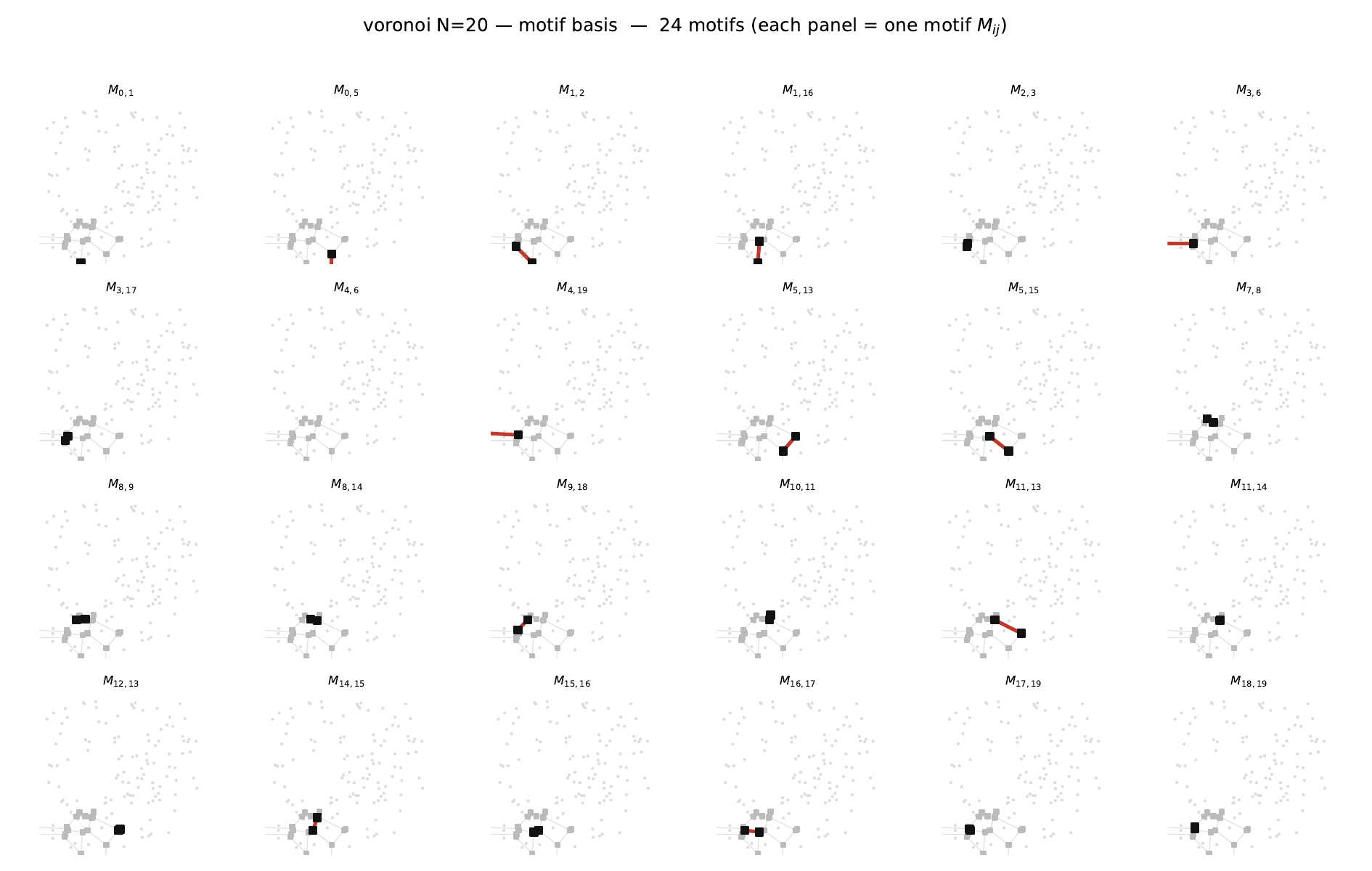}
\caption{Motif basis of the COSMOS Voronoi subgraph at $N=20$. Each
   panel is one motif $M_{ij}$, a single candidate edge shown in red,
   over a faint skeleton of the whole basis on the shared node layout.
   The basis carries one panel per co-visited pair, so at the sizes of
   Table~\ref{tab:compare} it runs to several hundred of them; a small
   graph is shown here because the construction, not its extent, is the
   point.}
\label{fig:motif-voronoi}
\end{figure*}

\begin{figure*}
\centering
\includegraphics[width=\textwidth]{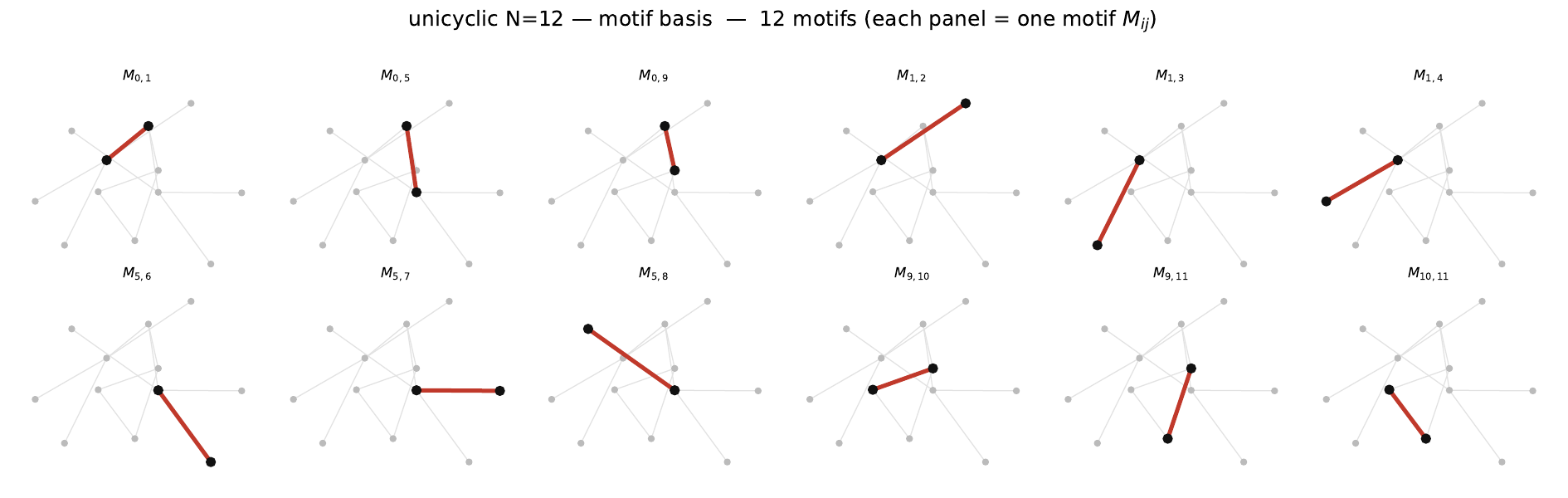}
\caption{Motif basis of the unicyclic control graph, the smallest case:
   twelve motifs for twelve edges. Conventions as in
   Fig.~\ref{fig:motif-voronoi}.}
\label{fig:motif-tree}
\end{figure*}

Figures~\ref{fig:motif-voronoi} and~\ref{fig:motif-tree} enumerate the
motif basis element by element for two of the test-beds: iterating over
the motifs (the support of the co-visitation,
Sect.~\ref{sec:forward}), each panel highlights a single motif
$M_{ij}$, one candidate edge, in red over a faint skeleton of the whole
basis, on the shared node layout. This makes explicit what the model is
built from: one pairwise motif per co-visited pair, whose log-weight
$\beta_{ij}$ is the only quantity the fit adjusts. For the clean
datasets used here the support is a subset of the true edges, so every
panel corresponds to a genuine candidate edge. The construction is the
same on every test-bed and the panel count simply follows $m$, so the
two smallest cases are shown and the rest omitted.

   \section{Discussion}
   \label{sec:discussion}

Across five graph families of different structural character (an
empirical email graph, two geometrically-embedded COSMOS graphs, and
two controlled graphs), the pipeline yields high-fidelity
reconstructions in both the multiplicative-noise regime, exactly and at
every size tested (Table~\ref{tab:mult}) and the finite-walk regime
(Table~\ref{tab:compare}), and does so up to full graph size. On the finite-walk data
(Table~\ref{tab:compare}) the COSMOS Delaunay and Voronoi graphs are
reconstructed at their full extent (MCC $0.988$ and $0.977$, no
subgraph boundary), and the empirical \texttt{email-Eu-core} graph
reaches MCC $0.967$ at $N=240$. On the full Delaunay graph the
graphical-lasso reference returns MCC $0.540$ against $0.988$ for
fbLM. The one place
the method shows strain is the densest empirical graph, where the
walk coverage becomes the binding constraint and vertices begin to be
missed altogether, a walk-coverage limitation rather than a failure of the
fit, and a natural target for the readout characterisation discussed
below.

The central practical conclusion concerns reconstructibility. In the
finite-walk regime the residual false negatives are, to an audit,
edges that the sampled walk never traverses; they carry no signal in
the data and cannot be recovered by any estimator from that sample.
The quality of a reconstruction is therefore set by how thoroughly
the walk covers the graph, and the cost curve of
Sect.~\ref{sec:res-cost} makes this explicit: a modest number of
walks already saturates the recoverable fraction. This separates two
questions that are easily conflated: how much the estimator can
extract, and how much the data contain, and locates the operative
limit in the sampling rather than in the fit.

Two directions follow naturally. The first is scale: the
polynomial per-iteration cost (a finite-difference Jacobian and a
pseudoinverse) leaves room to push beyond the sizes reported here,
provided walk coverage is maintained. The second is adversarial
robustness. When observed transitions can be mis-recorded
deliberately, as chaff, or through measurement error, the
empirical support is polluted with spurious pairs, and support
pollution rather than count noise becomes the dominant failure mode;
characterising the readout in that regime connects this
reconstruction problem to the security of distributed information
storage on networks \citep{zlatic2026}, where the question of how
much of a network an adversary can reconstruct from observed traffic
is central. We defer a quantitative treatment to future work.

\begin{appendix}

\section{The group-weight solver}
\label{app:stiefel}

This appendix states Eq.~\eqref{eq:stiefel} operationally, so that the
fit can be reproduced from the text alone.

\paragraph{Whitening.}
The per-vertex blocks $g^{(0)}_i=J^{(i)}(J^{(i)})^{\top}$ differ in
overall scale by orders of magnitude, since a high-degree vertex
contributes a much larger residual block than a leaf. They are therefore
whitened by the inverse square root of their sum before entering the
objective. With
\begin{equation}
   \bar g=\sum_{k} g^{(0)}_k+\epsilon I,
   \qquad
   \bar g = V\Lambda V^{\top},
   \qquad
   \bar g^{-1/2}=V\Lambda^{-1/2}V^{\top},
\end{equation}
the normalised blocks of Eq.~\eqref{eq:stiefel} are
\begin{equation}
   \tilde g^{(0)}_i=\bar g^{-1/2}\,g^{(0)}_i\,\bar g^{-1/2},
   \label{eq:whiten}
\end{equation}
with the eigenvalues of $\bar g$ clipped from below at $\epsilon$. The
whitening is what makes the objective scale-free: without it the
maximiser is dominated by the largest block and the weights collapse
onto the highest-degree vertex.

\paragraph{Optimisation.}
Let $A\in\mathbb{R}^{n\times m}$ have rows $a_i^{\top}$. The retraction
below returns orthonormal \emph{rows}, $AA^{\top}=I_n$, whenever
$n\le m$, which covers every test-bed reported here except the
radialness control ($n=12$, $m=11$), where the same retraction returns
orthonormal columns instead. Writing
$q_i(A)=a_i^{\top}\tilde g^{(0)}_i a_i+\epsilon$, the objective of
Eq.~\eqref{eq:stiefel} is $\sum_i\log q_i(A)$, whose Euclidean gradient
has rows
\begin{equation}
   (\nabla A)_i=\frac{2}{q_i(A)}\,\tilde g^{(0)}_i\,a_i .
\end{equation}
This is projected onto the tangent space of the manifold,
\begin{equation}
   \mathrm{grad}=\nabla A-A\,\mathrm{sym}\!\big(A^{\top}\nabla A\big),
   \qquad
   \mathrm{sym}(M)=\tfrac12(M+M^{\top}),
\end{equation}
and the iterate is retracted by the polar retraction, computed as
$A\leftarrow UV^{\top}$ from the thin singular value decomposition
$A+\alpha\,\mathrm{grad}=U\Sigma V^{\top}$. The projection above is
tangent to either manifold: when $AA^{\top}=I$ one has
$A\,\mathrm{sym}(A^{\top}G)A^{\top}=\mathrm{sym}(GA^{\top})$, so
$\mathrm{sym}(\mathrm{grad}\,A^{\top})=0$, and when $A^{\top}A=I$ the
condition $\mathrm{sym}(A^{\top}\mathrm{grad})=0$ follows directly. The
iteration starts from a Gaussian random $A$ orthonormalised the same
way, with a fixed seed.

\paragraph{Extraction of the weights.}
The weights are not multipliers but the value of the objective's
per-vertex term at the optimum,
\begin{equation}
   w_i=q_i(A_\star)=a_{i\star}^{\top}\,\tilde g^{(0)}_i\,a_{i\star}+\epsilon,
   \label{eq:wextract}
\end{equation}
so $w_i$ measures how much of the whitened information at vertex $i$ is
captured along the direction the frame assigns to it, which is the
reading set out in Sect.~\ref{sec:dynwlm}. Equation~\eqref{eq:wextract}
is the only place the frame $A$ enters the fit: it is recomputed at each
Levenberg--Marquardt iteration and discarded once the weights are read
off.

\paragraph{Constants.}
Table~\ref{tab:const} lists the numerical values of the finite-walk runs
of Table~\ref{tab:compare}, which share one setting across every
test-bed. The two auxiliary experiments were run with their own
iteration budgets, recorded here for completeness: the
multiplicative-noise rows of Table~\ref{tab:mult} use $25$ Stiefel
iterations at $N=25$ and $N=37$, and $80$ Levenberg--Marquardt
iterations at $N=12$; the coverage study of Fig.~\ref{fig:cost} uses
$35$ Levenberg--Marquardt and $25$ Stiefel iterations. Both fit a
complete pairwise basis, in which every vertex carries at least one
parameter, so the floor of Eq.~\eqref{eq:S} is unnecessary there and is
absent from those runs; the gauge invariance is correspondingly exact
for them. Nothing else differs, and no constant was tuned to a
particular graph.

\begin{table}[h]
\centering
\caption{Numerical constants of the finite-walk runs. The iteration
   budgets of the two auxiliary experiments differ and are given in the
   text.}
\label{tab:const}
\begin{tabular}{llc}
\toprule
Symbol & Meaning & Value \\
\midrule
$\beta^{(0)}$ & parameter initialisation & $0$ \\
$\lambda_0$ & initial damping & $10^{2}$ \\
& damping on accept / reject & $\lambda/2$ / $10\lambda$ \\
$\Delta$ & finite-difference step for $\mathbf J$ & $10^{-3}$ \\
& finite-difference step for $\mathbf J_{\rho}$ & $10^{-4}$ \\
& Levenberg--Marquardt iterations & $60$ \\
$\alpha$ & Stiefel step size & $0.1$ \\
& Stiefel iterations & $60$ \\
$\epsilon$ & ridge in Eqs.~\eqref{eq:whiten}, \eqref{eq:wextract} & $10^{-6}$ \\
& floor on off-diagonal $S$ & $10^{-12}$ \\
& cutoff in $(\Pi\mathcal I\Pi)^{+}$ & $10^{-8}$ \\
& group weights solved for & $m\le700$ \\
\bottomrule
\end{tabular}
\end{table}

\end{appendix}

\end{document}